\documentclass[12pt,a4paper]{article}

\usepackage{jheppub}

\usepackage{etoolbox}
    \makeatletter
    \patchcmd{\maketitle}{\@fpheader}{}{}{}
    \makeatother

\usepackage[utf8]{inputenc}
\usepackage{lmodern}
\usepackage{exscale}

\usepackage{amsmath}
\usepackage{amssymb}
\usepackage{mathtools}

\usepackage{mathrsfs}

\usepackage{tensor}

\usepackage[low-sup]{subdepth}

\newcommand{\scr}{\scriptscriptstyle}

\usepackage{graphicx}
\newcommand{\longsim}{\scalebox{1.8}[1]{$\sim$}}

\usepackage{xcolor}
\hypersetup{
    colorlinks,
    linkcolor={black},
    citecolor={black},
    urlcolor={black}
}

\usepackage[normalem]{ulem}

\usepackage{cancel}

\makeatletter
\newcommand{\dalembertian}{\mathop{\mathpalette\dalembertian@\relax}}
\newcommand{\dalembertian@}[2]{%
  \begingroup
  \sbox\z@{$\m@th#1\square$}%
  \dimen0=\fontdimen8
    \ifx#1\displaystyle\textfont\else
    \ifx#1\textstyle\textfont\else
    \ifx#1\scriptstyle\scriptfont\else
    \scriptscriptfont\fi\fi\fi3
  \makebox[\wd\z@]{%
    \hbox to \ht\z@{%
      \vrule width \dimen0
      \kern-\dimen0
      \vbox to \ht\z@{
        \hrule height \dimen0 width \ht\z@
        \vss
        \hrule height 2\dimen0
      }%
      \kern-2.5\dimen0
      \vrule width 2.5\dimen0
    }%
  }%
  \endgroup
}
\makeatother

\graphicspath{ {./tadpole_images/} }

\title{When tadpoles matter: One-loop corrections \\ 
for spectator Higgs in inflation}

\author[a]{Dra\v{z}en Glavan,}
\emailAdd{glavan@fzu.cz}
\author[b]{Tomislav Prokopec}
\emailAdd{t.prokopec@uu.nl}

\affiliation[a]{CEICO, FZU --- Institute of Physics of the Czech Academy of Sciences,
	\\
	Na Slovance 1999/2, 182 21 Prague 8, Czech Republic}
\affiliation[b]{Institute for Theoretical Physics, Spinoza Institute \& EMME$\Phi$,
	Utrecht University,
	\\Buys Ballot Building, Princetonplein 5,
	3584 CC Utrecht, The Netherlands}

\abstract{
We consider the classical attractor regime of the spectator Abelian Higgs model in power-law inflation,
and compute the one-loop corrections to its evolution. For computations we utilize dimensional
regularization and the propagators in the unitary gauge.
The corrections to both the scalar condensate
and the energy-momentum tensor exhibit secular ultraviolet contributions,
that tend to slow down the rolling of the scalar
down its potential, and drive it away from the classical attractor. These corrections need not be
suppressed if the~$U(1)$ charge is much larger than the scalar self-coupling, which is seen already in
flat space. In addition, at late times the secular corrections 
necessarily invalidate the perturbative loop expansion. We find the late time secular corrections to 
be captured by the renormalization group, which opens up the possibility to resum them past
the breakdown of perturbativity.

}

\notoc

\begin{document}

\maketitle

\titlepage

\section{Introduction}
\label{sec: Introduction}

The question of whether inflationary observables --- expressed in terms of scalar and tensor
cosmological perturbations --- can be significantly affected by quantum loop corrections is
still very much an open one. Quantum corrections to cosmological perturbations
can arise in different manners: (i) as indirect corrections descending from the quantum corrections
to the background fields, and (ii) as direct corrections arising from interactions between
evolving perturbations.
On a technical level, the indirect corrections descend from the corrections to the 
one-point functions (condensates), while the direct corrections are corrections to
the connected two-point functions. The computation of former type is simpler as
typically the loop consists of a single propagator.

While the understanding of the direct type of corrections to the cosmological
perturbations is still marred by questions of gauge 
dependence~\cite{Miao:2017feh},
considerable progress has been made in understanding the indirect type of corrections,
to which this work is devoted to as well. When it comes to corrections to condensates
in cosmological expanding and accelerating spaces
distinctions should be made between ultraviolet and infrared corrections.
The former are in a sense universal as they depend on the ultraviolet
structure of the theory captured by modes that are not strongly coupled to the background.
That is why ultraviolet corrections  in principle can be computed in any curved 
space~\cite{Birrell:1982ix,Elizalde:1993ee,Elizalde:1993qh}.
These have found applications in inflation, investigating quantum corrections to the
inflaton potential~\cite{Markkanen:2012rh,Herranen:2016xsy},
particularly in
Higgs inflation~\cite{Bezrukov:2007ep,Bezrukov:2013fka,Bezrukov:2014ipa}
in connection to the quantum corrections to the Higgs effective 
potential~\cite{Mooij:2011fi,George:2012xs,George:2013iia,George:2015nza,Fumagalli:2016lls,Fumagalli:2016sof},
and in studies of how curvature corrections influence the Higgs potential 
stability~\cite{DiLuzio:2014bua,Herranen:2014cua,Herranen:2015ima,Figueroa:2017slm,Markkanen:2018bfx,Mantziris:2022fuu} 
 in inflation and reheating
(for a review see~\cite{Markkanen:2018pdo}).

Infrared corrections derive from the tree-level effect of gravitational particle 
production~\cite{Parker:1968mv,Parker:1969au,Parker:1971pt}
for fields non-conformally coupled to gravity, such as scalars or gravitons. 
It should be emphasized that infrared corrections
can be more important that ultraviolet corrections. 
These effects are innate
to accelerating expanding spacetimes, and have no analogue in flat space, as they are in effect a 
consequence of the cosmological horizon.
As opposed to ultraviolet corrections, infrared ones are not universal and depend on the details of the 
model and the expanding spacetime, and it is much more challenging to quantify them. 
That is the reason why most works in the literature make a simplifying assumption 
of approximating the expanding spacetime by exactly de Sitter space, characterized by the constant Hubble rate.
Such works date back to~\cite{Birrell:1982ix,Shore:1979as,Allen:1983dg,Vilenkin:1982sg}, and the investigations of curvature 
corrections to symmetry breaking potentials and the study of phase transitions.
Early results on corrections to the inflaton evolution suggested that the corrections are tiny~\cite{Bilandzic:2007nb,Herranen:2013raa}, 
which is due to the smallness of the couplings
and the assumed perturbative regime.
However, strong infrared effects for spectator fields can lead to symmetry restoration of the scalar models with symmetry-breaking 
potentials~\cite{Starobinsky:1994bd,Serreau:2013eoa}, but also to dynamical symmetry breaking in massless scalar electrodynamics~\cite{Prokopec:2002uw,Prokopec:2002jn,Prokopec:2003iu}
and the generation of a non-perturbatively large mass for the vector 
field~\cite{Prokopec:2006ue,Prokopec:2007ak,Prokopec:2008gw}. 
When infrared effects are large typically one needs to describe them using non-perturbative methods,
the Starobinsky's stochastic formalism~\cite{Starobinsky:1986fx,Starobinsky:1994bd} being perhaps the most prominent one.

The de Sitter space is a very often good approximation for the realistic slow-roll inflationary background.
However, primordial cosmological observables crucially depend on the parameters measuring the deviation
of primordial inflation from the exact de Sitter space. This deviation is usually encoded in the slow-roll parameters,
where the principal one measures the rate of change of the Hubble rate,~$\epsilon\!=\!-\dot{H}/H^2$,
and the higher order ones encode the higher time derivatives. 
It was argued in Ref.~\cite{Miao:2015oba} that quantum corrections
introduce an additional fine-tuning problem into single scalar potential models of inflation,
and presumably the same is true for spectator fields.
This is why it is paramount to understand 
how quantum corrections depend on the slow-roll parameters. 
Efforts in that direction have been undertaken in inflation for spectator scalars~\cite{Janssen:2009pb,Prokopec:2015owa,Cho:2015pwa,Markkanen:2017edu}
and for the inflaton~\cite{Liao:2018sci,Miao:2019bnq,Miao:2020zeh,Kyriazis:2019xgj,Sivasankaran:2020dzp,Katuwal:2021kry},
and even for the period of reheating~\cite{Ema:2016dny,Katuwal:2022szw}.
Particularly insightful are studies of quantum effects in power-law inflation, 
characterized by constant principal slow-roll parameter~$\epsilon$, and subsequently 
promoted to an adiabatic function of time.
This is because analytically tractable computations are still feasible, both for the ultraviolet and for the infrared, and
one can get a clear picture of the effects that corrections can engender.
A particularly important reference for this work is~\cite{Janssen:2009pb} 
which
considered a symmetry-breaking scalar potential model
in power-law inflation with the assumption of the scaling solution,
and it found that inclusion of quantum effects 
can significantly affect the scalar potential by dynamically restoring the symmetry of the potential
in the small field regime, more rapidly than in de Sitter.

In this work we extend the analysis of Ref.~\cite{Janssen:2009pb} 
and compute one-loop corrections to the condensate of
the Abelian Higgs model --- scalar electrodynamics with a symmetry breaking potential.
For the sake of clarity we adopt several simplifying assumptions. We assume that the 
Abelian Higgs model is a spectator, in the sense that it does not source the evolution of the 
expanding background. For the background we assume it is exact power-law inflation,
characterized by the constant principal slow-roll parameter~$\epsilon$.
This still allows for an 
analytically tractable analysis of the problem, at least at one-loop order. We consider the scalar modulus
of the non-minimally coupled Abelian Higgs model to be in the attractor regime in power-law inflation,
where it tracks the evolution of the decreasing background Hubble rate,~$\overline{\phi} \!\propto\! H$.
This scaling attractor can be seen as a dynamical generalization of the symmetry breaking minimum 
in equilibrium theories, as it is formed by the competition between the non-minimal coupling
and the quartic self-coupling. 

The main focus of this work are quantum corrections to the
evolution of the scaling attractor.
We account for loop corrections to the attractor 
descending from both the charged scalar and the vector.
These are quantified by computing one-loop corrections to the scalar one-point function,
and to the graviton one-point function (that is sourced by the energy-momentum tensor.
Our analysis utilizes propagators for power-law inflation,
namely the scalar one thas was worked out in~\cite{Janssen:2008px},
and the vector one that was constructed in~\cite{Glavan:2020zne} in the unitary gauge.
An important difference between the analysis performed in this work 
and that in~\cite{Janssen:2009pb} is in that the only source 
of breaking of scaling we encounter 
descends from contributions connected to
the ultraviolet scale $\mu$
introduced by the counterterms.
The non-vanishing scalar condensate generates masses for scalar and vector 
perturbations, which regulates their infrared sector and prevents large 
infrared efects from developing, except in the small condensate limit
which is beyond the scope of this work.
Our finding show that there is a secular ultraviolet effect that typically
slows down the rolling of the spectator scalar down its potential,
and eventually drives the condensate away from the classical attractor.
Interestingly, this effect is completely absent in the de Sitter limit,
which again points to the importance of considering quantum effects in more
realistic inflationary backgrounds.

We perform the loop computations using dimensional regularization
and we adopt a rather conservative approach to renormalization.
Namely we forbid counterterm coefficients from
depending on the principal slow roll parameter $\epsilon$,
that we consider to be a geometric quantity,
that describes a particular realization of the system, and does not characterize the theory.
This severely restricts which terms can be subtracted by finite parts of the 
counterterms, thus making 
the results more robust to renormalization 
prescription dependencies.
Furthermore, we shy away from using any initial state subtractions,
as these introduce renormalization scheme dependence 
to the initial state, and generate corrections which anyway decay at late times we are interested in.

This work is organized as follows. In section~\ref{sec: Preliminaries}
we present the model and set the stage for the problem. Sections~\ref{sec: Scalar tadpole} 
and~\ref{sec: Energy-momentum tensor} are devoted to the calculation of the one-loop scalar 
condensate and energy-momentum tensor, where we discuss in detail
the countertems needed to renormalize the results in dimensional regularization.
Section~\ref{sec: Various limits} is devoted to studying
 various limits and comparing them to the literature.
Particular attention is paid to the analysis of secular effects
and their growth, and to the conditions under which 
pertubation theory breaks.
In section~\ref{sec: Late time limit and an RG explanation}
we present a preliminary analysis of how to 
emply renormalization group tools to restore perturbativity of the 
results. In particular, we show there 
that the RG resummation allows one to write 
the complete one-loop results in the tree-level form, up to
an additional term in the energy-momentum tensor, which is due to the perturbative nonrenormalizability of quantum gravity.
In section~\ref{sec: Discussion} we discuss our main results.

\section{Preliminaries}
\label{sec: Preliminaries}

The Abelian Higgs model is a commonly utilized as a toy model for 
the electroweak sector of the standard model. In this work we consider 
its nonminimally coupled variant, and the quantum corrections to its
dynamics in power-law inflation. In this section we summarize the properties 
of the background, and recount the main properties of the model.

\subsection{Power-law inflation}
\label{subsec: Power-law inflation}

Spatially flat expanding cosmological spaces are well described by the Friedmann-Lema\^{i}tre-Robertson-Walker (FLRW) invariant line element,
\begin{equation}
ds^2 = - dt^2 + a^2(t) d\vec{x}^{\,2}
	= a^2(\eta) \Bigl[ - d\eta^2 + d\vec{x}^{\,2} \Bigr] \, ,
\end{equation}
where time is conveniently given either in physical time coordinate~$t$ or the conformal time coordinate~$\eta$,
the two being related via the scale factor,~$dt \!=\! a d\eta$,
and where equal time spatial hypersurfaces are Euclidean spaces covered by Cartesian coordinates~$x^i$.
The expansion of the space is encoded by the scale factor~$a$, and its first several derivatives.
The first derivative is usually given as the physical Hubble rate~$H$,
 or the conformal Hubble rate~$\mathcal{H}$,
\begin{equation}
H = \frac{1}{a} \frac{da}{dt} \, ,
\qquad \qquad
\mathcal{H} = \frac{1}{a} \frac{d a}{ d\eta} \, ,
\qquad \qquad
\mathcal{H} = a H \, ,
\end{equation}
while in inflation the second derivative is encoded by the principal slow-roll parameter,
\begin{equation}
\epsilon 
	= - \frac{1}{H^2} \frac{d H}{ dt} 
	= 1 - \frac{1}{\mathcal{H}^2} \frac{d \mathcal{H}}{d\eta} \, .
\label{epsilon definition}
\end{equation}
 In this work we consider power-law inflation~\cite{Lucchin:1984yf,La:1989za}
that is defined by the constant slow-roll parameter,
\begin{equation}
\epsilon = {\tt const.} \, ,
\qquad \qquad
0< \epsilon  <  1 \, .
\label{constant epsilon}
\end{equation}
Even though this model is not a realistic model of inflation, as it is excluded by 
observations~\cite{Planck:2018vyg}
that favour
 an adiabatically evolving~$\epsilon$, 
it is more realistic than the de Sitter space,~$\epsilon\!=\!0$, often utilized to study quantum field theoretic
effects in inflation. In addition to being almost as mathematically tractable as de Sitter, power law inflation
incorporates the effect of the non-vanishing slow-roll parameter, and the evolving Hubble 
rate~$H\!=\!H_0 a^{-\epsilon}$.
The scale factor and the conformal Hubble rate in power-law inflation take the form,
\begin{equation}
a(\eta) = \Bigl[ 1 - (1\!-\!\epsilon) H_0(\eta\!-\!\eta_0) \Bigr]^{-\frac{1}{1-\epsilon}}
	\, ,
\qquad\quad
\mathcal{H}(\eta) = H_0 a^{1-\epsilon} \, ,
\end{equation}
where~$\eta_0$ is the initial time at which~$a(\eta_0) \!=\! 1$, and~$H_0 \!=\! \mathcal{H}(\eta_0)$.
The curvature tensors in power-law inflation are,
\begin{align}
R_{\mu\nu\rho\sigma}
	={}&
	2 H^2 g_{\mu [\rho} g_{\sigma] \nu}
	+ 4 \epsilon \bigl( a^2 \delta^0_{[\mu} g_{\nu][\sigma} \delta^0_{\rho]}  \bigr)
	\, ,
\\
R_{\mu\nu} ={}&
	(D \!-\! 1 \!-\! \epsilon) H^2 g_{\mu\nu}
	+ (D\!-\!2) \epsilon H^2 \bigl( a^2 \delta_\mu^0 \delta_\nu^0 \bigr) 
	\, ,
\\
R ={}& 
	(D\!-\!1)(D\!-\!2\epsilon) H^2 
\, .
\end{align}
In addition, it is useful to note that the Weyl tensor,
\begin{equation}
C_{\mu\nu\rho\sigma}
	\equiv
	R_{\mu\nu\rho\sigma}
	- \frac{4}{(D\!-\!2)} g_{\mu] [\rho} R_{\sigma] [\nu}
	+
	\frac{2 R}{ (D\!-\!1) (D\!-\!2) } g_{\mu[\rho} g_{\sigma]\nu}
	\, .
\label{Weyl tensor}
\end{equation}
%

\subsection{Nonminimally coupled Abelian Higgs model}
\label{subsec: Abelian Higgs model}


The Abelian Higgs model consists of an Abelian gauge field $A_\mu$ 
interacting with a charged complex scalar $\Phi$. Its action in $D$-dimensional curved space is,
\begin{align}
S \bigl[ A_\mu, \Phi, \Phi^* \bigr]
	={}& \int\! d^{D\!}x \, \sqrt{-g} \, \biggl[
		- \frac{Z_{\scr A}^0 }{4} g^{\mu\rho} g^{\nu\sigma} F_{\mu\nu} F_{\rho\sigma}
		- Z_\phi^0 g^{\mu\nu} \bigl( D_\mu \Phi \bigr)^{\!*} \bigl( D_\nu \Phi \bigr)
\nonumber \\
&	\hspace{3.5cm}
		- \lambda_0 \bigl( \Phi^* \Phi \bigr)^2
		- \xi_0 R \Phi^*\Phi
		\biggr]
		\, ,
\label{action SQED}
\end{align}
where $g_{\mu\nu}$ denotes the metric tensor [$g^{\mu\rho}g_{\mu\nu} \!=\! \delta^\rho_\nu$,
$g \!=\! {\rm det}(g_{\mu\nu})$],~$R$ is the Ricci scalar,
and the bare couplings (denoted by sub/superscripts 0) in~(\ref{action SQED}) 
are split into their renormalized values plus counterterms (denoted by prefix~$\delta)$,
\begin{equation}
Z_{\scr A}^0 = 1 + \delta Z_{\scr A} \, ,
\quad
Z_\phi^0 = 1 + \delta Z_\phi \, ,
\quad
\lambda_0 = \lambda + \delta\lambda \, ,
\quad
\xi_0 = \xi + \delta \xi \,,
\quad
q_0 = q + \delta q
\, .\quad
\label{bare and renormalized couplings}
\end{equation}
The counterterms are necessary to absorb ultraviolet divergences of quantum loops, and 
are organized as a power series in~$\hbar$, the dependence on which is henceforth suppressed 
by adopting the natural units~$\hbar\!=\!c\!=\!1$.
In the unitary gauge, defined by the condition~${\rm Im}(\Phi)\!=\!0$, the action~(\ref{action SQED}) 
reads~\cite{Glavan:2020zne},
\begin{align}
S[A_\mu, \phi] 
	={}&
	\int\! d^{D\!}x \, \sqrt{-g} \,
	\biggl[
	- \frac{Z_A^0}{4} g^{\mu\rho} g^{\nu\sigma} F_{\mu\nu} F_{\rho\sigma}
	- \frac{ (q_0 \phi)^2 }{2} g^{\mu\nu} A_\mu A_\nu
\nonumber \\
&	\hspace{3.5cm}
	- \frac{ Z_\phi^0 }{2} g^{\mu\nu} (\partial_\mu \phi) (\partial_\nu \phi)
	- \frac{\lambda_0}{4} \phi^4
	- \frac{\xi_0}{2} R \phi^2
	\biggr]
	\, .
\label{model}
\end{align}
where~$\phi\!\equiv\! {\rm Re}(\Phi)/\sqrt{2}$ is the canonically normalized real scalar field.
Note that in the limit~$q\!\to\!0$ the unitary gauge action reduces to the one for the self-interacting
real scalar field. This is the sense in which all the results presented here can be seen in the limit of 
vanishing charge.

Following~\cite{Glavan:2020zne}, we consider the Abelian Higgs model in rigid power-law inflation.
Only the scalar is supposed to have a homogeneous and isotropic condensate, 
which satisfies the equation of motion,
\begin{equation}
\Bigl[ \dalembertian - \xi R - \lambda \overline{\phi}^{\,2} \Bigr] \overline{\phi} = 0 \, ,
\label{tree-level condensate eom}
\end{equation}
that admits an attractor solution,
\begin{equation}
\overline{\phi} = \frac{\overline{\phi}_0}{H_0} H \, ,
\label{scalar condensate}
\end{equation}
with the amplitude set by the parameters of the model and the slow-roll parameter of the background,
\begin{equation}
\frac{\overline{\phi}_0}{H_0}
	=
	\pm \sqrt{ \frac{1}{\lambda} 
		\Bigl[ \epsilon (D \!-\! 1 \!-\! 2\epsilon) 
			\!-\! \xi (D\!-\!1) (D\!-\!2\epsilon) \Bigr] }
	\! \xrightarrow{D\to4}
	\pm \sqrt{ \frac{1}{\lambda} 
		\Bigl[ 
			(1 \!-\! 6 \xi)  (2\!-\!\epsilon) 
			\!-\! 2(1 \!-\! \epsilon)^2
			\Bigr] 
			}
			\, .
\label{scalar amplitude}
\end{equation}
For this attractor solution to exist the amplitude above should be real, which puts some bounds on the
non-minimal coupling,
\begin{equation}
\xi < \frac{\epsilon (3 \!-\! 2\epsilon)}{ 6 (2 \!-\! \epsilon) } 
	 \equiv \xi_{\rm cr} 
	\, ,
\label{inequality}
\end{equation}
dependent on the slow-roll parameter.
We study scalar and vector fluctuations around this background solution. The vector 
field~$A_\mu$ is assumed
not to have a classical condensate and is treated as a fluctuation, while the full scalar 
field,~$\phi\!=\! \overline{\phi} \!+\! \varphi$, is expanded in fluctuations~$\varphi$ around 
the classical attractor condensate defined in~(\ref{scalar condensate})
and~(\ref{scalar amplitude}).
We capture the nonlinear
evolution of quantized fluctuations~$\hat{\varphi}$ 
and~$\hat{A}_\mu$ perturbatively, by computing corrections to the evolution of
the linearized (non-interacting) fluctuations~$\hat{\varphi}^0$ and~$\hat{A}_\mu^0$.
We organize the corrections in powers of the linearized fields, which generates the 
usual loop expansion from the interaction picture.
Each power of the linearized field counts as one power of~$\sqrt{\hbar}$, which we henceforth suppress by adopting natural
units~$\hbar\!=\!1$.

Linearized perturbations around the attractor condensate~(\ref{scalar condensate}) and 
their two-point functions have been worked out in the unitary gauge in~\cite{Glavan:2020zne}.
The linearized scalar perturbations,
\begin{equation}
S_{\scr S}^{\scr (2)}[A_\mu, \varphi] 
	=
	\int\! d^{D\!}x \, \sqrt{-g} \,
	\biggl[
	- \frac{ 1 }{2} g^{\mu\nu} (\partial_\mu \varphi) (\partial_\nu \varphi)
	- \frac{3\lambda}{2} \overline{\phi}^{\,2} \varphi^2
	- \frac{\xi}{2} R \varphi^2
	\biggr]
	\, .
\label{scalar perturbations action}
\end{equation}
correspond to a spectator scalar with an effective mass~$ 3 \lambda \overline{\phi}^2$
and the nonminimal coupling to gravity~$\xi$. Since the tree-level condensate~(\ref{scalar condensate})
scales as the Hubble rate,
for the purposes of dynamics we can consider the linearized scalar perturbations to possess just an effective
mass,
\begin{equation}
M_{\scr S}^2\!=\! 3 \lambda \overline{\phi}^2 \!+ \xi R
	=
	\Bigl[ 3\epsilon (D \!-\! 1 \!-\! 2\epsilon) 
			\!-\! 2\xi (D\!-\!1) (D\!-\!2\epsilon) \Bigr] H^2
	\! \xrightarrow{D\to4} \!
	3 \Bigl[- 4\xi (2\!-\!\epsilon) + \epsilon (3\!-\!2\epsilon)  \Bigr] H^2
	,
\label{scalar mass}
\end{equation}
or an effective non-minimal coupling,
\begin{equation}
\xi_{\scr S} = \frac{M_{\scr S}^2 }{ R}
	\, \xrightarrow{D\to4} \,
	- 2\xi + \frac{\epsilon (3\!-\!2\epsilon) }{2(2\!-\!\epsilon)} 
	\, .
\end{equation}
Nevertheless, when computing the energy-momentum tensor, it is important to make the distinction between 
the two commensurate contributions in~(\ref{scalar perturbations action}). The latter of the two quantities
is bound by~$\xi_{\scr S} \!>\! (3\!-\!2\epsilon)\epsilon/[6(2\!-\!\epsilon)]$ due to~(\ref{inequality}),
which ensures there are no infrared divergences for CTBD states~\cite{Ford:1977in,Janssen:2009nz}.

Linearized vector perturbations,
\begin{equation}
S_{\scr V}^{\scr (2)}[A_\mu]
	= \int\! d^{D\!}x \, \sqrt{-g} \, \biggl[
		- \frac{1}{4} g^{\mu\rho} g^{\nu\sigma} F_{\mu\nu} F_{\rho\sigma}
		- \frac{ \bigl( q \overline{\phi} \bigr)^{\!2} }{2} g^{\mu\nu} \! A_\mu A_\nu
		\biggr]
\end{equation}
behave as though massive, with the effective mass induced by the scalar condensate,
\begin{equation}
M_{\scr V}^2 = \bigl( q \overline{\phi} \bigr)^{\!2}
	=
	\frac{q^2}{\lambda} 
		\Bigl[ \epsilon (D \!-\! 1 \!-\! 2\epsilon) 
			\!-\! \xi (D\!-\!1) (D\!-\!2\epsilon) \Bigr] H^2
	\! \xrightarrow{D\to4}
	\frac{q^2}{\lambda} 
		\Bigl[ 
			- 6 \xi (2\!-\!\epsilon) 
			\!+\! \epsilon ( 3 \!-\! 2\epsilon)
			\Bigr] H^2
			\, ,
\label{vector mass}
\end{equation}
or as effectively non-minimally coupled,
\begin{equation}
\xi_{\scr V} 
	= \frac{ M_{\scr V}^2 }{R}
	\xrightarrow{D\to4}
	\frac{q^2}{\lambda} 
		\biggl[ 
			- \xi 
			\!+\! \frac{ \epsilon ( 3 \!-\! 2\epsilon) }{ 6(2\!-\!\epsilon) }
			\biggr]
			\, .
\end{equation}
Eq.~(\ref{inequality}) implies $\xi_{\scr V} \!>0$ and $M_{\scr V}^2>0$,
which ensures stability and infrared finiteness of vector perturbations,
and distinguishes our model from vector curvaton models~\cite{Dimopoulos:2008rf}.

In this work we compute the one-loop effects that interacting perturbations impart on 
the evolution of the scalar condensate in the attractor~(\ref{scalar condensate}), and on 
the energy-momentum tensor;
the latter accounts for the backreaction onto the power-law inflation. This computation
requires the two-point functions of scalar and vector perturbations, and in particular 
their coincidence limits collected in the following subsection.

\subsection{Two-point functions}
\label{subsec: Two-point functions}

The two-point functions  for linearized fluctuations of the model 
in~(\ref{model}) in the unitary gauge in power-law inflation have 
been worked out in~\cite{Glavan:2020zne}. 
 They are expectation values of linearized fields (denoted by superscript 0).
In this work we compute the one-loop corrections to the scalar condensate, 
and to the energy-momentum tensor.
These are simplest one-loop corrections, where the loop is formed by a single propagator only.
That is why we require only the dimensionally regulated coincidence limits of 
the propagators from~\cite{Glavan:2020zne}.
For the scalar this is,
\begin{equation}
\bigl\langle \hat{\varphi}^0(x) \hat{\varphi}^0(x) \bigr\rangle 
	=
	\boldsymbol{\Gamma}(\nu_{\scr S})
		(1\!-\!\epsilon)^2 H^2 
		\biggl[ \Bigl( \frac{D \!-\! 3}{2} \Bigr)^{\!2} \!- \nu_{\scr S}^2 \biggr]
		\, ,
\label{scalar coincidence}
\end{equation}
and for the vector,
\begin{align}
\MoveEqLeft[3]
\bigl\langle \hat{A}_\mu^0(x) \hat{A}_\nu^0(x) \bigr\rangle 
	=
	\boldsymbol{\Gamma}(\nu_{\scr V})
	(1\!-\!\epsilon)^2 H^{2} 
	\biggl[
	\Bigl( \frac{D \!-\! 3}{2} \Bigr)^2 \! - \nu_{\scr V}^2
	\biggr]
\nonumber \\
&	\times \Biggl\{
	 \frac{1}{D} \biggl[
	 (D\!-\!1)
	 +
	 \biggl( D\!-\!1 \!+\! \frac{(D \!-\! 4)\epsilon}{2(1 \!-\! \epsilon)} \biggr) 
	 \biggl( 2 \!+\! \frac{(D\!-\!4)(2\!-\!\epsilon)}{2(1\!-\!\epsilon)} \biggr) 
	 \frac{(1\!-\!\epsilon)^2 H_0^2}{(q \overline{\phi}_0)^2} \biggr] g_{\mu\nu}
\nonumber \\
&	\hspace{1cm}
	+
	\frac{(D \!-\! 4)\epsilon}{2(1 \!-\! \epsilon)}
	\biggl( 2 \!+\! \frac{(D \!-\! 4)(2 \!-\! \epsilon)}{2(1 \!-\! \epsilon)} \biggr) 
		\frac{(1\!-\!\epsilon)^2 H_0^2}{(q\overline{\phi}_0)^2} 
			\bigl( a^2 \delta_\mu^0 \delta_\nu^0 \bigr)
	\Biggr\} 
	\, ,
\label{vector coincidence}
\end{align}
where we have defined a coefficient,
\begin{align}
\MoveEqLeft[1]
\boldsymbol{\Gamma}(\nu)
\equiv
	\frac{\Gamma\bigl( \frac{2-D}{2} \bigr) \, \bigl[ (1\!-\!\epsilon) H \bigr]^{D-4} }{ (4\pi)^{ \frac{D}{2} } } 
	\frac{ \Gamma \bigl( \frac{D-3}{2} \!+\! \nu \bigr) \,
					\Gamma\bigl( \frac{D-3}{2} \!-\! \nu \bigr) }
				{ \Gamma \bigl( \frac{1}{2} \!+\! \nu \bigr) \,
						\Gamma \bigl( \frac{1}{2} \!-\! \nu \bigr) }
\label{Gamma def}
\\
&
\overset{D\to4}{\longsim} \,
	\frac{\Gamma\bigl( \frac{2-D}{2} \bigr) \, \bigl[ (1\!-\!\epsilon) H \bigr]^{D-4} }{ (4\pi)^{ \frac{D}{2} } } 
	\Biggl\{ 1 + \frac{(D\!-\!4)}{2} \biggl[ 
		\psi\Bigl( \frac{1}{2} \!+\! \nu \Bigr)
		+ \psi\Bigl( \frac{1}{2} \!-\! \nu \Bigr) \biggr]
	+ \mathcal{O}\bigl[ (D\!-\!4)^2 \bigr]
	\Biggr\}
	\, ,
\nonumber 
\end{align}
divergent in~$D\!=\!4$ on the account 
of~$\Gamma\bigl( \frac{2-D}{2} \bigr) \, \overset{D\to4}{\longsim} \, 2/(D\!-\!4)$,
that depends on indices of scalar and vector perturbations, respectively,
\begin{align}
\nu_{\scr S}^2
	={}&
	\biggl( \frac{D \!-\! 1 \!-\! \epsilon}{ 2 (1\!-\!\epsilon) } \biggr)^{\!\!2}
		- \frac{\xi_{\scr  S} (D\!-\!1) (D\!-\!2\epsilon) }{ (1\!-\!\epsilon)^2 } 
	\xrightarrow{D\to4}
	\frac{25}{4}
		- \frac{ 2 (1 \!-\! 6\xi) (2 \!-\! \epsilon) }{ (1\!-\!\epsilon)^2 } 
		\, ,
\label{nuS}
\\
\nu_{\scr V}^2 ={}&
	\biggl( \frac{D \!-\! 3 \!-\! \epsilon}{ 2 (1\!-\!\epsilon) } \biggr)^{\!\!2}
		- \frac{\xi_{\scr  V} (D\!-\!1) (D\!-\!2\epsilon) }{ (1\!-\!\epsilon)^2 } 
	\xrightarrow{D\to4}
	\frac{1}{ 4 }
		+ \frac{2q^2}{ \lambda } \biggl[
			1 \!-\! \frac{ ( 1 \!-\! 6 \xi ) (2\!-\!\epsilon) }{ 2 (1\!-\!\epsilon)^2 }
			\biggr] 
	\, .
\label{nuV}
\end{align}
In four dimensions the range of these indices is limited by~(\ref{inequality}),
\begin{equation}
\nu_{\scr S}^2 \in \Bigl( -\infty, \frac{9}{4} \Bigr) \, ,
\qquad \qquad
\nu_{\scr V}^2 \in \Bigl( -\infty, \frac{1}{4} \Bigr) \, .
\end{equation}
We also need a dimensionally regulated scalar kinetic term,
which is easily computed by first taking derivatives of the scalar Wightman two-point function and 
then using expressions from Sec. 7.5.3 of~\cite{Glavan:2020zne},~\footnote{If the coincidence
limit is defined in terms of the Feynman propagator instead care needs to be taken to 
discard possible local terms resulting from derivatives acting on the time-ordeing.
In addition, in the unitary gauge attention must be paid to the fact that the Feynman propagator 
is not the Green's function of the theory~\cite{Glavan:2020zne}, and to the difference between~$\mathcal{T}$
 and~$\mathcal{T}^*$ ordered products, similarly to what was pointed out in perturbative quantum gravity 
 by Donoghue~\cite{Donoghue:2020hoh}. }
\begin{align}
\bigl\langle \partial_\mu \hat{\varphi}^0(x) \,
	\partial_\nu \hat{\varphi}^0(x) \bigr\rangle 
	={}&
	\boldsymbol{\Gamma}(\nu_{\scr S}) 
		(1\!-\!\epsilon)^4 H^4
		\biggl[ \Bigl( \frac{D \!-\! 3}{2} \Bigr)^{\!2} \!- \nu_{\scr S}^2 \biggr] 
\nonumber  \\
&	\times
	\Biggl\{
	- \frac{1}{D}
	\biggl[
	\Bigl( \frac{D \!-\! 1}{2} \Bigr)^{\!2} \!- \nu_{\scr S}^2
	\biggr]
	g_{\mu\nu}  
	+
	\biggl[ \frac{(D\!-\!2) \epsilon }{2 (1\!-\!\epsilon)} \biggr]^2 
		\bigl( a^2 \delta_\mu^0 \delta_\nu^0 \bigr)
	\Biggr\}
	\, .
\end{align}
Likewise we need the coincidence limit of the vector field strenght correlator, 
obtained by taking the coincident limit of the
correlator computed in Sec. 8 of~\cite{Glavan:2020zne},
\begin{align}
\MoveEqLeft[2]
\bigl\langle \hat{F}_{\mu\nu}^0(x) \hat{F}_{\rho\sigma}^0(x) \bigr\rangle 
	=
	\boldsymbol{\Gamma}(\nu_{\scr V}) 
	(1 \!-\! \epsilon)^4 H^4
		\biggl[ \Bigl( \frac{D \!-\! 3}{2} \Bigr)^{\!2} \!- \nu_{\scr V}^2 \biggr]
\\
&
	\times
	\Biggl\{
	- \frac{4}{D} 
	\biggl[ \Bigl( \frac{D \!-\! 1}{2} \Bigr)^{\!2} \!- \nu_{\scr V}^2 \biggr]
	g_{\mu [\rho} g_{\sigma] \nu}
	-
	\frac{2 (D\!-\!4)\epsilon}{(1 \!-\! \epsilon)} 
		\biggl[ 1 - \frac{(D \!-\! 4)\epsilon}{2(1 \!-\! \epsilon)} \biggr]
	\bigl( a^2 \delta_{[\mu}^0 g_{\nu] [\sigma } \delta_{\rho]}^0 \bigr)
	\Biggr\}
	\, .
\nonumber 
\end{align}

Even though for convenience we have introduced a number of parameters in this section,
it should be emphasized that there are only two independent coupling constants in the 
problem,~$\lambda$ and~$q$, and two background parameters,~$\epsilon$ and~$H_0$.
Nevertheless, we use the parameters introduced here interchangeably, to make the expressions
more compact. 
 Six particularly useful combinations when presenting final results are are,
\begin{subequations}%
\begin{align}
(1\!-\!\epsilon)^2 H^2 
	\xrightarrow{D\to4} {}&
	- \frac{\lambda \overline{\phi}^{\,2}}{2} 
	+ \frac{1}{2} \Bigl( \frac{1}{6} \!-\! \xi \Bigr) R
	\, ,
\label{useful1}
\\
(1\!-\!\epsilon)^2 \Bigl( \frac{1}{4} \!-\! \nu_{\scr S}^2 \Bigr) H^2
	\xrightarrow{D\to4} {}&
	3 \lambda \overline{\phi}^2 - \Bigl( \frac{1}{6} \!-\! \xi \Bigr) R
	\, ,
\label{useful2}
\\
(1\!-\!\epsilon)^2 \Bigl( \frac{9}{4} \!-\! \nu_{\scr S}^2 \Bigr) H^2
	\xrightarrow{D\to4} {}&
	2 \lambda \overline{\phi}^2
	\, ,
\label{useful3}
\\
(1\!-\!\epsilon)^2 \Bigl( \frac{25}{4} \!-\! \nu_{\scr S}^2 \Bigr) H^2
	\xrightarrow{D\to4} {}&
	2 \Bigl( \frac{1}{6} \!-\! \xi \Bigr) R
	\, ,
\label{useful4}
\\
(1\!-\!\epsilon)^2 \Bigl( \frac{1}{4} \!-\! \nu_{\scr V}^2 \Bigr) H^2
	\xrightarrow{D\to4} {}&
	\bigl( q \overline{\phi} \bigr)^{\!2}
	\, ,
\label{useful5}
\\
(1\!-\!\epsilon)^2 \Bigl( \frac{9}{4} \!-\! \nu_{\scr V}^2 \Bigr) H^2
	\xrightarrow{D\to4} {}&
	\bigl( q \overline{\phi} \bigr)^{\!2}
	- \lambda \overline{\phi}^2 + \Bigl( \frac{1}{6} \!-\! \xi \Bigr) R
	\, .
\label{useful6}
\end{align}
\label{useful}%
\end{subequations}
%

\section{Scalar tadpole}
\label{sec: Scalar tadpole}

\begin{figure}[h]
\centering
%
%
%
%
%
%
\hspace{1cm}
\includegraphics[width=4cm]{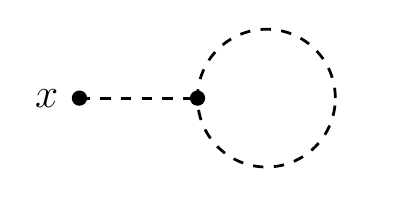}
\hfill
\includegraphics[width=4cm]{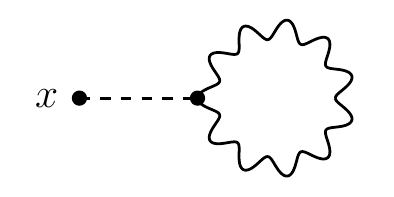}
\hfill
\includegraphics[width=3cm]{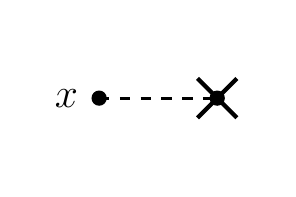}
\hspace{1cm}
\vskip-4mm
\caption{\linespread{1}\selectfont 
Diagrams depicting the one-loop corrections to the one-point function from 
the scalar perturbation.
Dashed line correspond to the scalar propagator, and wavy ones to the vector propagator.
The first two diagrams descend from the vertices defined in the cubic action~(\ref{cubic action}),
which also defines the counterterms represented by the last diagram.}
\label{tadpole diagrams}
\end{figure}

Expanding the unitary gauge action~(\ref{model}) to cubic order in perturbations,
\begin{equation}
S^{\scr (3)}[A_\mu, \varphi] 
	= \!\!
	\int\! \! d^{D\!}x \, \sqrt{-g} \,
	\biggl[
	- \overline{\phi} \Bigl(q^2 \varphi A^\mu A_\mu
		- \lambda \varphi^3
		\Bigr)
	- \varphi \Bigl( 
		\delta \lambda \overline{\phi}^{\,3}
		+ \delta\xi R \overline{\phi}
		- \delta Z_\phi \dalembertian \overline{\phi} 
		\Bigr)
	\biggr]
	\, ,
\label{cubic action}
\end{equation}
defines the vertices and counterterms that generate the one-loop corrections to the 
tadpole, whose diagramatic representation is given in Fig.~\ref{tadpole diagrams}.
Note that the divergent counterterm coefficients count as two powers of pertrubations.

The equation of motion for the one-loop correction to the scalar one-point function takes the form,
\begin{equation}
\Bigl[ \dalembertian - \xi R - 3 \lambda \overline{\phi}^{\,2} \Bigr]
	\bigl\langle \hat{\varphi} \bigr\rangle
	=
	\mathcal{S}_{\scr S} + \mathcal{S}_{\scr V} + \delta \mathcal{S}
	\, ,
\label{tadpole eom}
\end{equation}
Where the scalar loop and the vector loop sources, corresponding to the first and the second diagram
in Fig.~\ref{tadpole diagrams} respectively, are
\begin{equation}
\mathcal{S}_{\scr S} =
	3 \lambda \overline{\phi} \, \bigl\langle \hat{\varphi}^0 \hat{\varphi}^0 \bigr\rangle 
	\, ,
\qquad \qquad
\mathcal{S}_{\scr V} =
	q^2 \overline{\phi} \, g^{\mu\nu} \bigl\langle \hat{A}_\mu^0 \hat{A}_\nu^0 \bigr\rangle 
	\, ,
\label{S+V source}
\end{equation}
and the counterterms, corresponding to the last amputated diagram in Fig.~\ref{tadpole diagrams}
 contribute as,
\begin{align}
\delta \mathcal{S}
	={}&
	\bigl( \delta \xi \!-\! \xi \delta Z_\phi \bigr) R \overline{\phi}
	+ \bigl( \delta \lambda \!-\! \lambda \delta Z_\phi \bigr) \overline{\phi}^{\,3}
\nonumber \\
={}&
	\biggl[ \bigl( \delta \xi \!-\! \xi \delta Z_\phi \bigr) (D\!-\!1)(D\!-\!2\epsilon) H^2 
	+ \bigl( \delta \lambda \!-\! \lambda \delta Z_\phi \bigr) \overline{\phi}^{\,2}
	\biggr] \overline{\phi}
	\, ,
\label{deltaS}
\end{align}
where we have used the tree-level equation of motion~(\ref{tree-level condensate eom}) 
to write them in this form. The counterterm coefficients will be chosen so as to
absorb the divergences from the scalar and the vector source~(\ref{S+V source}).
These divergences are absorbed independently for each source, up to arbitrary finite parts.
Accordingly, we split the counterterm contribution to the source into three parts,
\begin{equation}
\delta \mathcal{S} 
	= 
	\bigl[ \delta \mathcal{S} \bigr]_{\scr S}^{\rm div.}
	+
	\bigl[ \delta \mathcal{S} \bigr]_{\scr V}^{\rm div.}
	+
	\bigl[ \delta \mathcal{S} \bigr]^{\rm fin.}
	\, ,
\end{equation}
and this split translates onto the analogous split of all the counterterm coefficients.
Since the renormalizability of the theory should not depend on the background spacetime,
the counterterm coefficients ahould be independent of the packground parameters such 
as~$\epsilon$. This breaks some redundancy between the coefficients.

\subsection{Scalar loop source}
\label{subsec: Scalar loop source}

This is the simplest part to compute. 
Scalar tadpole source can be written as, using the coincident scalar two-point 
function~(\ref{scalar coincidence}) and expressions from Sec.~\ref{subsec: Abelian Higgs model},
\begin{align}
\mathcal{S}_{\scr S}
={}&
	\overline{\phi} \,
	\boldsymbol{\Gamma}(\nu_{\scr S})
	\Biggl\{
		\frac{3 D (D\!-\!2)(D \!-\! 6) \lambda }{32(D \!-\! 1)}
		\biggl[ 1 \!-\! \frac{4(D \!-\! 1) \xi}{ (D \!-\! 2) } \biggr] 
		(D\!-\!1) (D\!-\!2\epsilon) H^2
\nonumber \\
&	\hspace{4cm}
	+
	\frac{ 3 (D \!+\! 2) (8 \!-\! D) \lambda^2 }{8}
	\overline{\phi}^{\,2}
	+
	\mathcal{O}\bigl[ (D\!-\!4)^2 \bigr]
	\Biggr\} 
	\, .
\end{align}
Comparing with the the counterterms~(\ref{deltaS}),
the simplest~$\epsilon$-independent choices are,
\begin{subequations}
\begin{align}
\bigl[ \delta \xi \!-\! \xi \delta Z_\phi \bigr]_{\scr S}^{\rm div.}
	={}& 
	\frac{ \mu^{D-4} \, \Gamma \bigl( \frac{2-D}{2} \bigr) }{ (4\pi)^{ \frac{D}{2} } }
		\times 
		\frac{3 D (D\!-\!2) (6 \!-\! D) \lambda }{32(D \!-\! 1)}
		\biggl[ 1 \!-\! \frac{4(D \!-\! 1) \xi}{ (D \!-\! 2) } \biggr]
		 \, ,
\\
\bigl[ \delta \lambda \!-\! \lambda \delta Z_\phi \bigr]_{\scr S}^{\rm div.}
	={}& 
	\frac{ \mu^{D-4} \, \Gamma \bigl( \frac{2-D}{2} \bigr) }{ (4\pi)^{ \frac{D}{2} } }
		\times \frac{3(D\!+\!2) (D\!-\!8) \lambda^2 }{8} 
		\, .
\end{align}
\label{scalar tadpole counterterms}%
\end{subequations}
Taking the~$D\!\to\!4$ limit of the scalar loop tadpole source produces a fully renormalized result,
\begin{equation}
\mathcal{S}_{\scr S} + \bigl[ \delta \mathcal{S} \bigr]_{\scr S}^{\rm div.}
	\xrightarrow{D\to4}
	\frac{ 3 \lambda \overline{\phi} H^2 }{ 16 \pi^2 }
	(1\!-\!\epsilon)^2 \Bigl( \frac{1}{4} \!-\! \nu_{\scr S}^2 \Bigr) 
	\biggl[
		2\ln \Bigl[ \frac{ (1\!-\!\epsilon) H }{ \mu } \Bigr]
		+
		\Psi(\nu_{\scr S})
	\biggr] 
		\, ,
\label{renrom scalar source}
\end{equation}
where we defined,
\begin{equation}
\Psi(\nu) \equiv
		\psi\Bigl( \frac{1}{2} \!+\! \nu \Bigr)
		+
		\psi\Bigl( \frac{1}{2} \!-\! \nu \Bigr) 
		\, ,
\label{Psi def}
\end{equation}
where~$\psi(z) \!=\! \frac{{d}}{{d}z}\ln[\Gamma(z)]$ denotes the digamma function.

\subsection{Vector loop source}
\label{subsec: Vector loop source}

For the vector tadpole source we only need the fully contracted coincident vector 
propagator from~(\ref{vector coincidence}). Using some relations between parameters 
from Sec.~(\ref{subsec: Abelian Higgs model}) we write it in the form,
\begin{align}
\mathcal{S}_{\scr V} ={}&
	\overline{\phi} \,
	\boldsymbol{\Gamma}(\nu_{\scr V})
	\Biggl\{
	\frac{(D\!-\!1) q^2 }{2} 
	\Bigl[2 q^2  \!-\! ( D \!-\! 2 ) \lambda \Bigr] 
	\overline{\phi}^{\,2}
	+
	3(D \!-\! 4) q^2
	\biggl[ 
		\frac{ 3 }{4}
		-  \frac{ (1\!-\!\epsilon)^3 H_0^2}{(q \overline{\phi}_0)^2} \biggr]
	\epsilon H^2 
\nonumber \\
&	\hspace{-0.2cm}
	+
	\frac{ ( D \!-\! 2 ) (8 \!-\! D) q^2 }{16}
	\biggl[ 1 
		\!-\! \frac{8(D\!-\!1) \xi}{( 8 \!-\! D )} \biggr]
		(D\!-\!1)(D\!-\!2\epsilon) H^2
	+ 
	\mathcal{O} \bigl[ (D\!-\!4)^2 \bigr]
		\Biggr\}
	\, .
\end{align}
Comparing this source with the form of the counterterms it is clear that the two divergent terms
have the same structure. Therefore, the simplest~$\epsilon$-independent choice for counterterm 
coefficients seems to be,
\begin{align}
\bigl[ \delta \xi \!-\! \xi \delta Z_\phi \bigr]_{\scr V}^{\rm div.} 
	={}&
	\frac{ \mu^{D-4} \, \Gamma\bigl( \frac{2-D}{2} \bigr) }{ (4\pi)^{ \frac{D}{2} } }
	\times
	\frac{ ( D \!-\! 2 ) ( D \!-\! 8 ) q^2 }{16}
	\biggl[ 1 \!-\! \frac{8(D\!-\!1) \xi}{(8 \!-\! D)} \biggr]
	\, ,
\\
\bigl[ \delta \lambda \!-\! \lambda \delta Z_\phi \bigr]_{\scr V}^{\rm div.} 
	={}&
	\frac{ \mu^{D-4} \, \Gamma\bigl( \frac{2-D}{2} \bigr) }{ (4\pi)^{ \frac{D}{2} } }
	\times
	\frac{(D\!-\!1) q^2 }{2} 
	\Bigl[ ( D \!-\! 2 ) \lambda \!-\! 2 q^2 \Bigr] 
	\, .
\end{align}
This leads to the renormalized vector tadpole source,
\begin{align}
\mathcal{S}_{\scr V} 
	+ \bigl[ \delta \mathcal{S} \bigr]_{\scr V}^{\rm div.}
\xrightarrow{D\to4} {}&
	\frac{ 3 q^2 \overline{\phi} H^2 }{ 16 \pi^2 } 
	\biggl\{
	(1\!-\!\epsilon)^2 \Bigl( \frac{9}{4} \!-\! \nu_{\scr V}^2 \Bigr)
	\biggl[
		2\ln \Bigl[ \frac{ (1\!-\!\epsilon) H }{ \mu } \Bigr]
		+ \Psi(\nu_{\scr V})
	\biggr]
\nonumber \\
&	\hspace{2.5cm}
	+
	\biggl[ 
		\frac{ 3 }{2}
		-  \frac{ 2(1\!-\!\epsilon)^3 H_0^2}{ \bigl( q \overline{\phi}_0 \bigr)^{\!2} } \biggr]
	\epsilon
	\biggr\}
	\, .
\label{renrom vector source}
\end{align}
%

\subsection{Solving tadpole equation}
\label{subsec: Solving tadpole equation}

Here we solve equation~(\ref{tadpole eom}) for the condensate 
correction~$\bigl\langle \hat{\varphi} \bigr\rangle$,
with sources computed in~(\ref{renrom scalar source})
and~(\ref{renrom vector source}). We can write the equation as,
\begin{align}
- \biggl[ \frac{1}{a^2} \partial_0^2 
	+ \frac{2H}{a} \partial_0 
	+ 6 \xi_{\scr S} (2\!-\!\epsilon) H^2 \biggr] 
	\bigl\langle \hat{\varphi} \bigr\rangle
={}&
	\frac{ 3 \overline{\phi} H^2 }{16 \pi^2}
	\biggl[ 2 \mathcal{A} \ln\Bigl[ \frac{(1\!-\!\epsilon) H}{\mu} \Bigr] + \mathcal{B} \biggr]
	\, ,
\label{tadpole eom parametrized}
\end{align}
where the coefficients are read off from the sources~(\ref{renrom scalar source})
and~(\ref{renrom vector source}) computed in the preceding sections.
The first coefficient,~$\mathcal{A}\!=\!\mathcal{A}_{\scr S} \!+\! \mathcal{A}_{\scr V}$, 
receives contributions from both the scalar and vector loop source, which are, respectively, succinctly
written as,
\begin{equation}
\mathcal{A}_{\scr S} =
	\lambda (1\!-\!\epsilon)^2 \Bigl( \frac{1}{4} \!-\! \nu_{\scr S}^2 \Bigr)
		\, ,
\qquad \qquad
\mathcal{A}_{\scr V} =
	q^2 (1\!-\!\epsilon)^2 \Bigl( \frac{9}{4} \!-\! \nu_{\scr V}^2 \Bigr)
	\, .
\end{equation}
Analogous is true for the second
coefficient,~$\mathcal{B} \!=\! \mathcal{B}_{\scr S} \!+\! \mathcal{B}_{\scr V} \!+\! \delta\mathcal{B}$, 
where scalar and vector contributions are, respectively,
\begin{subequations}
\begin{align}
\mathcal{B}_{\scr S} ={}&
	\lambda (1\!-\!\epsilon)^2 \Bigl( \frac{1}{4} \!-\! \nu_{\scr S}^2 \Bigr) \Psi(\nu_{\scr S})
	\, ,
\\
\mathcal{B}_{\scr V} ={}&
	q^2 (1\!-\!\epsilon)^2 \Bigl( \frac{9}{4} \!-\! \nu_{\scr V}^2 \Bigr)
	\Psi(\nu_{\scr V})
	+
	q^2 \epsilon
	\biggl[ \frac{3 }{2} - \frac{2(1\!-\!\epsilon)^3 H_0^2}{ \bigl( q \overline{\phi}_0 \bigr)^{\!2} } \biggr]
	\, ,
\end{align}
\end{subequations}
but in addition it receives a contribution from the finite parts of the counterterms,
\begin{equation}
\frac{3 \, \delta\mathcal{B} }{16 \pi^2}  =
	6(2 \!-\! \epsilon)  \bigl[ \delta \xi \!-\! \xi \delta Z_\phi \bigr]^{\rm fin.} 
	+ 
	\frac{ \overline{\phi}_0^{\,2} }{H_0^2} 
	\bigl[ \delta \lambda \!-\! \lambda \delta Z_\phi \bigr]^{\rm fin.} \, .
\label{deltaB}
\end{equation}
The solution of equation~(\ref{tadpole eom parametrized}) is not complicated,
\begin{equation}
\bigl\langle \hat{\varphi} \bigr\rangle
	=
	\frac{3 \overline{\phi} }{16 \pi^2} 
	\biggl[ 
	2 \mathscr{A} \ln\Bigl[ \frac{(1\!-\!\epsilon) H}{\mu} \Bigr]
	+ \mathscr{B}
	\biggr]
	+
	\frac{3 \overline{\phi} }{16 \pi^2} 
	\biggl[ 
	\mathscr{C}_{\scr +} a^{p_+} 
	+ \mathscr{C}_{\scr -} a^{p_-}
	\biggr]
	\, ,
\label{condensate correction}
\end{equation}
where the left term is a dynamical contribution,
with the two coefficients fixed by the sources of the equation of motion~(\ref{tadpole eom parametrized}),
\begin{equation}
\mathscr{A} =
	- \frac{H_0^2 }{ 2 \lambda \overline{\phi}_0^{\,2} } \mathcal{A} 
	\, ,
\qquad \qquad
\mathscr{B} =
	-\frac{H_0^2 }{ 2 \lambda \overline{\phi}_0^{\,2} } 
		\Bigl[ \mathcal{B} - 6 \epsilon (1\!-\!\epsilon)\mathscr{A} \Bigr]
		\, ,
\end{equation}
while the right term contains two free constants of integration
corresponding to homogeneous solutions that redshift away, as their powers,
\begin{equation}
p_{\scr \pm} =
	- (1 \!-\! \epsilon) \Bigl( \frac{3}{2} \pm \nu_{\scr S} \Bigr) 
	\, ,
\end{equation}
both have a negative real part (note that~$\nu_{\scr S}$ can be real or imaginary).
Imposing initial conditions on~$\bigl\langle \hat{\varphi} \bigr\rangle$ and its derivative
at the initial time~$\eta_0$ fixes the free integration constants~$\mathscr{C}_{\scr +}$
and~$\mathscr{C}_{\scr -}$ in~(\ref{condensate correction}).  
Requiring that the condensate corrections are minimized at the initial moment~$\eta_0$,
in the sense that both $\bigl\langle \hat{\varphi} \bigr\rangle(\eta_0) \!=\! 0$
and~$\partial_0 \bigl\langle \hat{\varphi} \bigr\rangle (\eta_0) \!=\!0$, 
fixes the two constants,~\footnote{The choice of 
$\mathscr{C}_{\scr \pm}$ in~(\ref{initial state subtractions}) is valid both when 
$p_\pm$ are real or a pair of complex conjugate numbers. In the latter case, 
$(\mathscr{C}_{\scr +}a^{p_+})^* \!=\! \mathscr{C}_{\scr -}a^{p_-}$ such that $\langle \hat\varphi\rangle$ 
in~(\ref{condensate correction}) remains real.}
\begin{subequations}
\begin{align}
\mathscr{C}_{\scr +} ={}&
	\frac{ p_{\scr -} }{ (p_{\scr +} \! - p_{\scr -}) }
	\biggl[ 
	2 \mathscr{A} \ln\Bigl[ \frac{(1\!-\!\epsilon) H_0}{\mu} \Bigr]
	+ \mathscr{B}
	+
	\frac{ 2 \epsilon \mathscr{A} }{p_{\scr -}}
	\biggr]
	\, ,
\\
\mathscr{C}_{\scr -} ={}&
	\frac{ p_{\scr +} }{ (p_{\scr -}\! - p_{\scr +}) }
	\biggl[ 
	2 \mathscr{A} \ln\Bigl[ \frac{(1\!-\!\epsilon) H_0}{\mu} \Bigr]
	+ \mathscr{B}
	+
	\frac{ 2 \epsilon \mathscr{A} }{p_{\scr +}}
	\biggr]
	\, .
\end{align}
\label{initial state subtractions}%
\end{subequations}
The choice of such coefficients amounts to a finite renormalization of the initial state,
and is important when discussing the fine-tuning issues in
inflation~\cite{Miao:2015oba}.
However, here we are primarily interested in
corrections descending from dynamics. Given that the state-dependent contributions decay in time in comparison, we shall henceforth not consider them,
and set~$\mathscr{C}_{\scr +}\!=\!\mathscr{C}_{\scr -}\!=\!0$. 
Thus, the one-loop correction to
the condensate receives additive contributions,~$\bigl\langle \hat{\varphi} \bigr\rangle \!=\! \bigl\langle \hat{\varphi} \bigr\rangle_{\scr S} \!+\! \bigl\langle \hat{\varphi} \bigr\rangle_{\scr V}$,
from the scalar and the vector loop, respectively,
\begin{align}
\bigl\langle \hat{\varphi} \bigr\rangle_{\scr S} ={}&
	\! - 
	\frac{3 \lambda \overline{\phi} }{16 \pi^2} 
	\biggl[ \frac{ 3 \lambda \overline{\phi}^2 \!-\! \bigl( \frac{1}{6} \!-\! \xi \bigr) R }
		{ 2 \lambda \overline{\phi}^{\,2} } \biggr]
	\biggl[
		2 \ln\Bigl[ \frac{(1\!-\!\epsilon) H}{\mu} \Bigr]
		+
		\Psi(\nu_{\scr S})
		+
		\frac{ 3 \epsilon (1\!-\!\epsilon) H^2 }{ \lambda \overline{\phi}^{\,2} } 
	\biggr]
	\, ,
\label{varphiS final}
\\
\bigl\langle \hat{\varphi} \bigr\rangle_{\scr V} ={}&
	- \frac{3 q^2 \overline{\phi} }{16 \pi^2} 
	\Biggl\{
	\biggl[ \frac{ \bigl( q \overline{\phi} \bigr)^{\!2} \!-\! \lambda \overline{\phi}^2 
		\!+\! \bigl( \frac{1}{6} \!-\! \xi \bigr) R}{ 2 \lambda \overline{\phi}^{2} }
		\biggr]
	\biggl[
		2 \ln\Bigl[ \frac{(1\!-\!\epsilon) H}{\mu} \Bigr]
	+
	\Psi(\nu_{\scr V})
\nonumber \\
&	\hspace{2cm}
	+ 
	\frac{ 3 \epsilon (1\!-\!\epsilon) H^2 }{ \lambda \overline{\phi}^{\,2} } 
	\biggr]
	+
	\frac{3 \epsilon H^2 }{4 \lambda \overline{\phi}^{\,2} } 
	\Biggr\}
	+
	\frac{3 \lambda \overline{\phi} }{16 \pi^2} 
	\!\times\!
	\frac{\epsilon (1\!-\!\epsilon)^3 H^4 }{ \bigl( \lambda \overline{\phi}^{\,2} \bigr)^{\!2} }
	\, ,
\label{varphiV final}
\end{align}
where the simplest choice for the finite parts of the counterterms in~(\ref{deltaB}) is~$\delta\mathcal{B}\!=\!0$,
which implies,
\begin{equation}
\bigl[ \delta\xi \!-\! \xi \delta Z_\phi \bigr]^{\rm fin.} = 0 \, ,
\qquad \quad
\bigl[ \delta \lambda \!-\! \lambda \delta Z_\phi \bigr]^{\rm fin.} = 0 \, ,
\label{two finite counters}
\end{equation}
that we henceforth assume. 
Note that in expressions above the argument of the logarithms, and the mode function indices
are expressed in terms of the scalar field and curvature scalar using~(\ref{useful}).

\section{Energy-momentum tensor}
\label{sec: Energy-momentum tensor}

In order to have a well defined one-loop energy-momentum tensor,
in addition to the action for the  model in~(\ref{model}), we need to consider
purely geometrical higher-derivative counterterms~\cite{tHooft:1974toh},
\begin{equation}
S_{\rm ctm.}[g_{\mu\nu}]
	= \int\! d^{D\!} x \, \sqrt{-g} \,
	\Bigl[
	\alpha_{\scr R} R^2 
	+
	\alpha_{\scr C} \mathcal{C}^2
	+
	\alpha_{\scr G} \mathcal{G}
	\Bigr]
	\, ,
\end{equation}
where the first counterterm is the square of the Ricci tensor, the second is the
square of the Weyl 
tensor~(\ref{Weyl tensor}),~$C^2 \!=\! C^{\mu\nu\rho\sigma} C_{\mu\nu\rho\sigma}$,
and the last one is the Gauss-Bonnet invariant~$\mathcal{G} \!=\! R^{\mu\nu\rho\sigma} R_{\mu\nu\rho\sigma} \!-\! 4 R^{\mu\nu} R_{\mu\nu} \!+\! R^2$.
Only the first of these counterterms contributes to renormalization on FLRW background.
The variation of the~$\mathcal{C}^2$ term is proportional to the Weyl tensor which vanishes 
for conformally flat spacetimes~(\ref{Weyl tensor}), 
and the Gauss-Bonnet part is only a surface term that does
not contribute in the bulk.
Thus, the full energy-momentum tensor operator is defined as
\begin{align}
\hat{T}_{\mu\nu} 
	={}& \frac{-2}{\sqrt{-g}} \frac{\delta S_*}{\delta g_{\mu\nu}} 
		\biggr|_{\substack{\phi\to\hat{\phi} \\  A\to \hat{A} } }
	=
	Z_A\biggl[ \delta_{(\mu}^\rho \delta_{\nu)}^\sigma 
		\!-\! \frac{1}{4} g_{\mu\nu} g^{\rho\sigma} \biggr] g^{\alpha\beta}
			\hat{F}_{\rho\alpha} \hat{F}_{\sigma\beta} 
\label{EMT operator}
 \\
&	
	+ \bigl( q_0\hat{\phi} \bigr)^2
		\biggl[ \delta_{(\mu}^\rho \delta_{\nu)}^\sigma
			\!-\! \frac{1}{2} g_{\mu\nu} g^{\rho\sigma} \biggr]
				\hat{A}_\rho \hat{A}_\sigma
	+ Z_\phi \biggl[ \delta_{(\mu}^\rho \delta_{\nu)}^\sigma
			\!-\! \frac{1}{2} g_{\mu\nu} g^{\rho\sigma} \biggr]
				\bigl( \partial_{\rho} \hat{\phi} \bigr) \bigl( \partial_{\sigma} \hat{\phi} \bigr)
\nonumber \\
&	
	- \frac{\lambda_0}{4} \hat{\phi}^{\,4} g_{\mu\nu}
	+ \xi_0 \Bigl[ G_{\mu\nu} + g_{\mu\nu} \dalembertian - \nabla_{\!\mu} \nabla_{\!\nu} \Bigr] 
		\hat{\phi}^{\,2} 
	+ \alpha_{\scr R} H_{\mu\nu}^{\scr R} \, ,
\nonumber
\end{align}
where the last term is,
\begin{align}
H_{\mu\nu}^{\scr R}
	={}& 
	4 \nabla_\mu \nabla_\nu R
	- 4 g_{\mu\nu} \dalembertian R
	- 4 R R_{\mu\nu}
	+ g_{\mu\nu} R^2  
\nonumber \\
&	\hspace{-0.5cm}
	\xrightarrow{\epsilon = {\rm const.}}
	(D\!-\!1) (D\!-\!2\epsilon) (D\!-\!4 \!-\! 6\epsilon) H^4 
	\Bigl[
	(D\!-\!1\!-\!4\epsilon) g_{\mu\nu}
	-
	4 \epsilon \bigl( a^2 \delta_\mu^0 \delta_\nu^0 \bigr)
	\Bigr]
	 \, .
\end{align}
We expand the energy-momentum tensor operator~(\ref{EMT operator}) up to quadratic order in fluctuating fields,
\begin{equation}
\hat{T}_{\mu\nu} = \overline{T}_{\mu\nu} + \hat{t}_{\mu\nu}^{\scr \, (1)} + \hat{t}_{\mu\nu}^{\scr \, (2)} + \dots
\end{equation}
where higher orders do not contribute at one-loop level. The three parts are defined and 
their expectation values computed in
the following three subsections.

\subsection{Tree-level part}
\label{subsec: Tree-level part}

The classical contribution to the energy-momentum tensor derives from the
scalar condensate~(\ref{scalar condensate}) only,
\begin{align}
\overline{T}_{\mu\nu}
	={}&
	\biggl[ \delta_\mu^\rho \delta_\nu^\sigma
			\!-\! \frac{1}{2} g_{\mu\nu} g^{\rho\sigma} \biggr]
				\bigl( \partial_\rho \overline{\phi} \bigr) \bigl( \partial_\sigma \overline{\phi} \bigr)
	- \frac{\lambda}{4} \overline{\phi}^{\,4} g_{\mu\nu}
	+ \xi \Bigl[ G_{\mu\nu} + g_{\mu\nu} \dalembertian - \nabla_{\!\mu} \nabla_{\!\nu} \Bigr] 
		\overline{\phi}^{\,2} 
		\, .
\end{align}
Evaluated on the attractor solution~(\ref{scalar condensate}) it reads,
\begin{align}
\overline{T}_{\mu\nu}
	={}&
	\frac{ (1\!-\!6\xi) \epsilon + (D\!-\!4)\xi }{4}
	\Bigl[
	- (D \!-\! 1 \!-\! 4\epsilon) g_{\mu\nu} + 4 \epsilon \bigl( a^2 \delta_\mu^0 \delta_\nu^0 \bigr)
	\Bigr]
	\bigl( \overline{\phi} H \bigr)^{\!2}
\nonumber \\
&
	\xrightarrow{D\to4}
	\frac{ (1\!-\!6\xi) \epsilon}{4}
	\biggl[
	- ( 3 \!-\! 4\epsilon ) g_{\mu\nu}
	+ 4\epsilon \bigl( a^2 \delta_\mu^0 \delta_\nu^0 \bigr)
	\biggr]
	\bigl( \overline{\phi} H \bigr)^{\!2}
	\, ,
\label{scalar classical Tmn 0}
\end{align}
and it is indeed covariantly conserved,~$ \nabla^\mu \overline{T}_{\mu\nu} \!=\! 0$.
It describes an effective ideal fluid with an equation of state dependent on the background,
\begin{equation}
\overline{w}
= -1 + \frac{4\epsilon}{3} \, .
\label{scaling of tree level Tmn}
\end{equation}
The energy-momentum tensor of the condensate, which is a spectator
from the perspective of the expansion, redshifts away faster
($\rho \!\propto\! H^4 \!\propto\! a^{-4\epsilon}$) than 
some fluid driving the expansion of the power-law inflation ($\rho\propto H^2 \propto a^{-2\epsilon}$)

\subsection{Part linear in fluctuations}
\label{subsec: Part linear in fluctuations}

Graviton couples to linearized fluctuations too.
This would be the contribution to the energy-momentum tensor coming from the 
correction to the scalar tadpole. However, this split should only go so far on the account
of it not being conserved by itself! Therefore, contributions cannot be split in a physical way.
\begin{figure}[h]
\centering
%
%
%
%
%
%
\includegraphics[width=5cm]{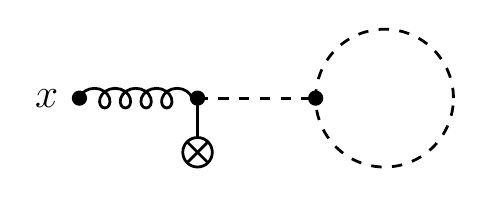}
\hfill
\includegraphics[width=5cm]{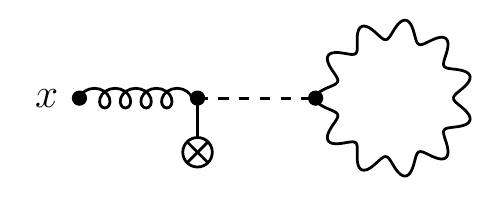}
\hfill
\includegraphics[width=4cm]{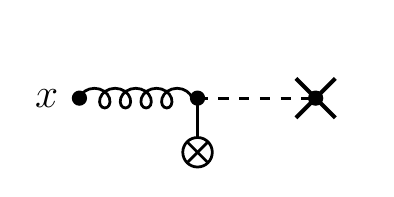}
\\
\vspace{-3mm}
\caption{\linespread{1}\selectfont 
Diagrams depicting the one-loop corrections to the graviton one-point function
descending from the one-loop corrections to the scalar one-point function in Fig.~\ref{tadpole diagrams}.
Curly lines correspond to the graviton propagator,
dashed line to the scalar propagator, and wavy lines to the vector propagator,
while encircled crosses stand for the classical condensate insertions.
The last diagram stands for counterterms. Amputating the graviton propagator leaves the
contribution to the one-loop energy-momentum tensor.}
\label{graviton tadpole 1}
\end{figure}
%
\begin{equation}
\bigl\langle \hat{t}_{\mu\nu}^{\,\scr (1)} \bigr\rangle
	= \Bigl[ 2\delta_{(\mu}^\rho \delta_{\nu)}^\sigma
			\!-\! g_{\mu\nu} g^{\rho\sigma} \Bigr]
				\bigl( \partial_\rho \overline{\phi} \bigr) 
				\partial_\sigma \bigl\langle \hat{\varphi} \bigr\rangle 
	- \lambda \overline{\phi}^{\,3} \bigl\langle \hat{\varphi} \bigr\rangle g_{\mu\nu}
	+ 2\xi \Bigl[ G_{\mu\nu} \!+\! g_{\mu\nu}\! \dalembertian - \nabla_{\mu} \nabla_{\nu}  \Bigr]
		\overline{\phi} \bigl\langle \hat{\varphi} \bigr\rangle 
		\, .
\label{T(1) general}
\end{equation}
This contribution is not covariantly conserved by itself. In fact, from the tree-level
equation of motion~(\ref{tree-level condensate eom}), and the one-loop tadpole 
equation~(\ref{tadpole eom}) it follows,
\begin{equation}
\nabla^\mu \bigl\langle \hat{t}_{\mu\nu}^{\,\scr (1)} \bigr\rangle
	=
	\bigl( \partial_\nu \overline{\phi} \bigr)
	\Bigl[
		\mathcal{S}_{\scr S} + \mathcal{S}_{\scr V} + \delta \mathcal{S}
	\Bigr]
	\, .
\label{T(1) non-conservation}
\end{equation}
This non-conservation has nothing to do with quantum anomalies, or quantum physics for that matter. 
Rather, it is a consequence of the part quadratic in fluctuations, discussed in the 
following section, contributing to the energy-momentum tensor at the same order as the 
part linear in fluctuations discussed in this section.

Plugging in the tree-level attractor solution for the condensate~(\ref{scalar condensate}),
and the one-loop condensate correction~(\ref{condensate correction}) into ~(\ref{T(1) general}) 
gives the tadpole contribution to the one-loop energy-momentum tensor.
Just like the condensate correction it receives dynamically generated contributions,
and the initial state dependent contributions,

\begin{equation}
	\bigl\langle \hat{t}_{\mu\nu}^{\,\scr (1)} \bigr\rangle 
	=
	\bigl\langle \hat{t}_{\mu\nu}^{\,\scr (1)} \bigr\rangle_{\rm dyn.}
	+
	\mathcal{T}_{\mu\nu}^{\scr +}
	+
	\mathcal{T}_{\mu\nu}^{\scr -}
	\, .
\end{equation}
where the former contribution is,
\begin{align}
\bigl\langle \hat{t}_{\mu\nu}^{\,\scr (1)} \bigr\rangle_{\rm dyn.} 
	={}&
	\frac{3 }{8 \pi^2}
	\biggl[ 2 \mathscr{A} \ln\Bigl[ \frac{(1\!-\!\epsilon)H}{\mu} \Bigr] + \mathscr{B} \biggr]
	\biggl[ \overline{T}_{\mu\nu}
		- \frac{\lambda}{4} \overline{\phi}^{\,4} g_{\mu\nu}
		\biggr] 
\label{scalar contribution to Tmn 1 ren}
\\
&	\hspace{-2cm}
	-
	\frac{ \bigl( \overline{\phi} H \bigr)^{\!2} \epsilon \mathscr{A} }{8 \pi^2}
	\biggl\{
		\Bigl[ (1 \!-\! 6\xi ) ( 2 \!-\! 5 \epsilon ) \!-\! 2( 1 \!-\! \epsilon ) \Bigr] g_{\mu\nu} 
		-
		\Bigl[ ( 1 \!-\! 6 \xi ) ( 1 \!+\! 5 \epsilon ) \!-\! ( 1 \!-\! \epsilon ) \Bigr]
			\bigl( a^2 \delta_\mu^0 \delta_\nu^0 \bigr) 
			\biggr\}
			 ,
\nonumber 
\end{align}
while the latter contributions are,
\begin{align}
\mathcal{T}_{\mu\nu}^{\scr \pm}
	={}&
	\frac{3 \mathscr{C}_{\scr \pm} a^{p_{\pm}}  \bigl( \overline{\phi} H \bigr)^{\! 2} }{16 \pi^2}
	\biggl\{
	g_{\mu\nu} \!\times\! 
			\bigl( 3 \!-\! 3\epsilon \!+\! p_{\scr \pm} \bigr)
			\Bigl[ 2 \xi \bigl( 1 \!+\! 2\epsilon \!-\! p_{\scr \pm} \bigr) \!-\! \epsilon \, \Bigr]
\nonumber \\
&	\hspace{1cm}
	+
	\bigl( a^2 \delta_\mu^0 \delta_\nu^0 \bigr) \!\times\! 
		2 \Bigl[
		\epsilon \bigl( \epsilon \!-\! p_{\scr \pm} \bigr)
		+ \xi p_{\scr \pm}
		- \xi \bigl( 3\epsilon \!-\! p_{\scr \pm} \bigr) \bigl( 2\epsilon \!-\! p_{\scr \pm} \bigr)
		\Bigr]
	\biggr\}
	\, .
\end{align}
These initial state dependent contributions are 
conserved by themselves,~$\nabla^\mu \mathcal{T}_{\mu\nu}^{\scr \pm} \!=\! 0$.
 They redshift away compared to the dynamical contribution and we do not consider them further.

\subsection{Part quadratic in fluctuations}
\label{subsec: Part quadratic in fluctuations}

\begin{figure}[h]
\centering
\hspace{1cm}
\includegraphics[width=4cm]{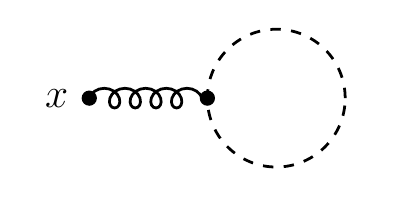}
\hfill
\includegraphics[width=4cm]{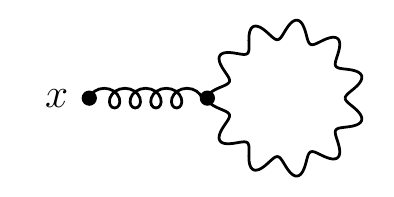}
\hfill
\includegraphics[width=3cm]{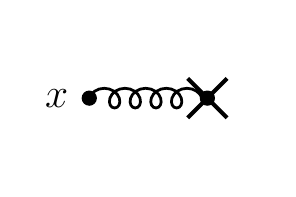}
\hspace{1cm}
\vspace{-3mm}
\caption{\linespread{1}\selectfont 
Diagrams depicting the one-loop corrections to the graviton one-point function 
descending from vacuum fluctuations.
Curly lines correspond to the graviton propagator,
dashed line to the scalar propagator, and wavy lines to the vector propagator.
The last diagram stands for counterterms. Amputating the graviton propagator leaves the
contribution to the one-loop energy-momentum tensor.}
\label{graviton tadpole 2}
\end{figure}
The renormalized expectation value of the quadratic part of the energy-momentum tensor
is a sum of all three contributions from Fig.~\ref{graviton tadpole 2},
\begin{equation}
\bigl\langle \hat{t}_{\mu\nu}^{\,\scr (2)} \bigr\rangle
	=
	\bigl\langle \hat{t}_{\mu\nu}^{\,\scr (2)} \bigr\rangle_{\scr S}
	+
	\bigl\langle \hat{t}_{\mu\nu}^{\,\scr (2)} \bigr\rangle_{\scr S}
	+
	\delta T_{\mu\nu}
	\, ,
\end{equation}
where the first contribution comes from the leftmost digram with a scalar loop,
\begin{align}
\bigl\langle \hat{t}_{\mu\nu}^{\,\scr (2)} \bigr\rangle_{\scr S}
	={}&
	\biggl[ \delta_{(\mu}^\rho \delta_{\nu)}^\sigma
			\!-\! \frac{1}{2} g_{\mu\nu} g^{\rho\sigma} \biggr]
				\bigl\langle \partial_\rho \hat{\varphi}^0 \,
				\partial_\sigma \hat{\varphi}^0 \bigr\rangle
	- 
	\frac{3\lambda}{2} \overline{\phi}^{\,2} \bigl\langle \hat{\varphi}^0 \hat{\varphi}^0 \bigr\rangle g_{\mu\nu}	
\nonumber \\
&
	+ 
	\xi \Bigl[ G_{\mu\nu} \!+\! g_{\mu\nu} \dalembertian -  \nabla_{\!\mu} \nabla_{\!\nu} \Bigr]
		\bigl\langle \hat{\varphi}^0 \hat{\varphi}^0 \bigr\rangle
		\, ,
\end{align}
the second contribution comes from the middle diagram with  the vector loop,
\begin{equation}
\bigl\langle \hat{t}_{\mu\nu}^{\,\scr (2)} \bigr\rangle_{\scr V} 
	=
	\biggl[ \delta_{(\mu}^\rho \delta_{\nu)}^\sigma g^{\alpha\beta}
		\!-\! \frac{1}{4} g_{\mu\nu} g^{\rho\sigma} g^{\alpha\beta} \biggr]
			\bigl\langle \hat{F}_{\rho\alpha}^0 \hat{F}_{\sigma\beta}^0 \bigr\rangle
	+ \bigl( q \overline{\phi} \bigr)^2
		\biggl[ \delta_{(\mu}^\rho \delta_{\nu)}^\sigma
			\!-\! \frac{1}{2} g_{\mu\nu} g^{\rho\sigma} \biggr]
				 \bigl\langle \hat{A}_\rho^0 \hat{A}_\sigma^0 \bigr\rangle
				 \, ,
\label{quadratic TV}
\end{equation}
and the remaining contribution comes from counterterms, denoted by the rightmost diagram,
\begin{align}
\delta T_{\mu\nu}
	={}&
	\delta Z_\phi \biggl[ \delta_{(\mu}^\rho \delta_{\nu)}^\sigma
			\!-\! \frac{1}{2} g_{\mu\nu} g^{\rho\sigma} \biggr] 
				\bigl( \partial_\rho \overline{\phi} \bigr) \bigl( \partial_\sigma \overline{\phi} \bigr)
	- 
	\frac{ \delta \lambda }{4} \overline{\phi}^{\,4} g_{\mu\nu}
\nonumber \\
&	\hspace{1.5cm}
	+ 
	\delta\xi
	\Bigl[ G_{\mu\nu} + g_{\mu\nu} \dalembertian - \nabla_{\!\mu} \nabla_{\!\nu} \Bigr]
		\overline{\phi}^{\,2} 
	+ 
	\delta \alpha_{\scr R} H_{\mu\nu}^{\scr R}
	\, .
\end{align}
This contribution is also not conserved on its own,
\begin{equation}
\nabla^\mu \bigl\langle \hat{t}_{\mu\nu}^{\,\scr (2)} \bigr\rangle 
	=
	-
	\bigl( \partial_\nu \overline{\phi} \bigr)
	\Bigl[
		\mathcal{S}_{\scr S} + \mathcal{S}_{\scr V} + \delta \mathcal{S}
	\Bigr]
	\, ,
\label{T(2) non-conservation}
\end{equation}
but rather its non-conservation precisely cancels the one in~(\ref{T(1) non-conservation}),
so that the entire one-loop energy momentum tensor is covariantly conserved, as it should be,
\begin{equation}
\nabla^\mu \Bigl( 
	\bigl\langle \hat{t}_{\mu\nu}^{\,\scr (1)} \bigr\rangle
	+
	\bigl\langle \hat{t}_{\mu\nu}^{\,\scr (2)} \bigr\rangle 
	\Bigr) = 0
	\, .
\end{equation}

The counterterm part, evaluated for power-law inflation reads,
\begin{align}
\delta T_{\mu\nu} ={}&
	g_{\mu\nu} \!\times\!
	\biggl\{
	- \frac{\delta\lambda}{4} \overline{\phi}^4
	+ \frac{\epsilon^2 }{2} \delta Z_\phi \bigl( \overline{\phi} H \bigr)^{\!2}
	+
	\delta \alpha_{\scr R} (D\!-\!1) (D\!-\!2\epsilon)
	(D\!-\!1\!-\!4\epsilon) (D \!-\! 4 \!-\! 6\epsilon) H^4
\nonumber \\
&	\hspace{1.cm}
	- 
	\frac{\delta\xi}{2} \Bigl[ (D\!-\!1)(D\!-\!2)
	- 6(D\!-\!2)\epsilon
	+ 12\epsilon^2
	\Bigr]
		H^2 \overline{\phi}^{\,2} 
	\biggr\}
\nonumber \\
&
	+
	\bigl( a^2 \delta_\mu^0 \delta_\nu^0 \bigr) \!\times\!
	\biggl\{
	\bigl( \delta Z_\phi \!-\! 6 \delta \xi \bigr) \epsilon^2 \bigl( \overline{\phi} H \bigr)^{\!2}
	+ 24 \delta \alpha_{\scr R} (D\!-\!1)(D\!-\!2\epsilon) \epsilon^2 H^4
\nonumber \\
&	\hspace{2.cm}
	+ (D\!-\!4) \delta \xi \epsilon \bigl( \overline{\phi} H \bigr)^{\!2}
	- 4 (D\!-\!4) \delta \alpha_{\scr R} (D\!-\!1) (D\!-\!2\epsilon) \epsilon H^4
	\biggr\}
	\, .
\label{T counterterms}
\end{align}
Just as for the tadpole, the scalar and the vector loops are renormalized independently
in the energy-monetum tensor, and we split the contributions,
\begin{equation}
\delta T_{\mu\nu}
	=
	\bigl[ \delta T_{\mu\nu} \bigr]_{\scr S}^{\rm div.}
	+
	\bigl[ \delta T_{\mu\nu} \bigr]_{\scr V}^{\rm div.}
	+
	\bigl[ \delta T_{\mu\nu} \bigr]^{\rm fin.}
	\, .
\end{equation}
%

\subsubsection{Scalar loop contribution}
\label{subsubsec: Scalar loop contribution}

For the scalar part the scalar parts of counterterms from before are enough. This is because
the vector parts of counterterms vanish for~$q \!\to\!0$.
\begin{align}
\MoveEqLeft[1]
\bigl\langle \hat{t}^{\scr \, (2)}_{\mu\nu} \bigr\rangle_{\scr S}
	= g_{\mu\nu} \times 
	\boldsymbol{\Gamma}(\nu_{\scr S}) (1\!-\!\epsilon)^2 H^2
		\biggl[ \Bigl( \frac{D \!-\! 3}{2} \Bigr)^{\!2} \!- \nu_{\scr S}^2 \biggr] 
 \\
&	\hspace{1cm}
	\times
		\frac{1}{D} \Biggl\{
		- \frac{(D\!-\!2)^2}{4} \biggl[ 1 - \frac{4(D \!-\! 1)\xi}{(D \!-\! 2)} \biggr] 
			( D\!-\!1 \!-\! D\epsilon ) \epsilon H^2 
		- 3 \lambda \overline{\phi}^{\,2}
		\Biggr\}
\nonumber \\
&	
	+ \bigl( a^2 \delta_\mu^0 \delta_\nu^0 \bigr) \times
		\boldsymbol{\Gamma}(\nu_{\scr S}) 
		\biggl[ 1 \!-\! \frac{ 4 (D \!-\! 1) \xi }{ (D \!-\! 2) } \biggr]
		\frac{(D\!-\!2)^2 \epsilon^2 H^2 }{ 32 }
		\times
		\Biggl\{ 
		(D \!+\! 2)  (8 \!-\! D) \lambda \overline{\phi}^{\,2 }
\nonumber \\
&	\hspace{1.5cm}
	-
	\frac{D (D\!-\!2) (6 \!-\! D)}{4 }
		\biggl[ 1 \!-\! \frac{4(D \!-\! 1) \xi}{(D \!-\! 2)} \biggr]
		 (D\!-\!2\epsilon)  H^2
	+
	\mathcal{O} \bigl[ (D \!-\! 4)^2 \bigr]
	\Biggr\}
	\, .
\nonumber
\end{align}
Since the factor~$\boldsymbol{\Gamma}(\nu_{\scr S})$ defined in~(\ref{Gamma def})
is divergent in~$D\!=\!4$, so are most of the contributions to the naive expectation 
value above. These need to be absorbed by the counterterms in~(\ref{T counterterms}),
by judiciously choosing four counterterm coefficients. Note that we already have two
conditions on these coefficients~(\ref{scalar tadpole counterterms}) from renormalizing 
the scalar contribution to the tadpole. Therefore, there are only two independent conditions
that we can require here. They are most conveniently found by requiring that the divergences
of the~$\bigl( a^2 \delta_\mu^0 \delta_\nu^0 \bigr)$ part are absorbed, since that is accomplished 
by a simple comparison with~(\ref{T counterterms}), and the simplest choice is
\begin{align}
\bigl[ \delta Z_\phi \!-\! 6 \delta \xi \bigr]_{\scr S}^{\rm div.} 
	={}&
		\frac{ \mu^{D-4} \, \Gamma \bigl( \frac{2-D}{2} \bigr) }{ (4\pi)^{ \frac{D}{2} } }
		\!\times\!
	\frac{ (D\!-\!2)^2(D \!+\! 2) (D \!-\! 8) \lambda }{32} 
		\biggl[ 1 \!-\! \frac{ 4 (D \!-\! 1) \xi }{ (D \!-\! 2) } \biggr]
		\, ,
\label{scalar counterterms for Tmn A}
\\
\bigl[ \delta \alpha_{\scr R} \bigr]_{\scr S}^{\rm div.} 
	={}&
	\frac{ \mu^{D-4} \, \Gamma \bigl( \frac{2-D}{2} \bigr) }{ (4\pi)^{ \frac{D}{2} } }
		\times
	\frac{D (D \!-\! 2)^2 }{ 768 (D\!-\!1)}
		\biggl[ 1 \!-\! \frac{ 4(D\!-\!1) \xi }{ (D \!-\! 2) } \biggr]^2
		\, .
\label{scalar counterterms for Tmn B}
\end{align}
In addition to fixing~$\delta \alpha_{\scr R}$, this fixes the three counterterms,
\begin{subequations}
\begin{align}
\bigl[ \delta \xi \bigr]_{\scr S}^{\rm div.} 
	={}&
	\frac{ \mu^{D-4} \, \Gamma \bigl( \frac{2-D}{2} \bigr) }{ (4\pi)^{ \frac{D}{2} } }
		\times 
		\frac{3 D \lambda \bigl[ 1 \!-\! (D\!-\!1)(D\!-\!2)\xi \bigr] }{8(D \!-\! 1)}
		\, ,
\label{scalar counterterm: xi}
\\
\bigl[ \delta \lambda \bigr]_{\scr S}^{\rm div.} 
	={}&
	\frac{ \mu^{D-4} \, \Gamma \bigl( \frac{2-D}{2} \bigr) }{ (4\pi)^{ \frac{D}{2} } }
		\times
		\biggl[ - \frac{9D\lambda^2}{4} \biggr]
		\, ,
\label{scalar counterterm: lambda}
\\
\bigl[ \delta Z_\phi \bigr]_{\scr S}^{\rm div.} 
	={}&
	\frac{ \mu^{D-4} \, \Gamma \bigl( \frac{2-D}{2} \bigr) }{ (4\pi)^{ \frac{D}{2} } }
		\times
		\Bigl[ - 3(D\!-\!4) \lambda \Bigr]
		\, ,
\label{scalar counterterm: Zphi}
\end{align}
\end{subequations}
which agrees (up to finite contributions) with the counterterms found in~\cite{Janssen:2009pb}.
Note that the wavefunction renormalization is finite, which agrees with what was also found in 
Ref.~\cite{Canevarolo:2022dvh} in a different setting.
Now we have for the renormalized scalar part of the energy-momentum tensor,
\begin{align}
\MoveEqLeft[1]
\bigl\langle \hat{t}^{\scr \, (2)}_{\mu\nu} \bigr\rangle_{\scr S}
	\!+\!
	\bigl[ \delta T_{\mu\nu} \bigr]_{\scr S}^{\rm div.}
	\! \xrightarrow{D\to4}
	- g_{\mu\nu} \!\times\! 
		\frac{ 3 \lambda \overline{\phi}^{\,2} }{64 \pi^2} (1\!-\!\epsilon)^2 H^2
		\Bigl( \frac{1}{4} \!-\! \nu_{\scr S}^2 \Bigr)
			\biggl[ 2 \ln\Bigl[ \frac{(1\!-\!\epsilon)H}{\mu} \Bigr] 
			\!+ \Psi(\nu_{\scr S}) 
			\!- \frac{1}{2}\biggr]
\nonumber \\
&	
	-
	\frac{ (1 \!-\! 6\xi) H^4 }{64 \pi^2}
		(1\!-\!\epsilon)^2 \Bigl( \frac{1}{4} \!-\! \nu_{\scr S}^2 \Bigr)
		\Biggl\{
			g_{\mu\nu} 
			\!\times \!
			\biggl[ (3 \!-\! 4\epsilon) \epsilon  \biggl( 2 \ln\Bigl[ \frac{(1\!-\!\epsilon)H}{\mu} \Bigr] + \Psi(\nu_{\scr S}) \biggr)
			\!+\! 1 \!-\! \frac{17\epsilon}{6}
			\biggr]
\nonumber \\
&	\hspace{3cm}
	-
		\bigl( a^2 \delta_\mu^0 \delta_\nu^0 \bigr) \!\times\! 4\epsilon
		\biggl[
		\epsilon \biggl( 2\ln\Bigl[ \frac{(1\!-\!\epsilon)H}{\mu} \Bigr] 
			+ \Psi(\nu_{\scr S})  \biggr)
		+ \frac{1}{3}
		\biggr]
		\Biggr\}
		\, .
\label{scalar contribution Tmn 2 ren}
\end{align}
The covariant contribution in the first line accounts for the non-conservation, while
the remainder is conserved by itself.

\subsubsection{Vector loop contribution}
\label{subsubsec: Vector loop contribution}

\begin{align}
\bigl\langle \hat{t}^{\scr \, (2)}_{\mu\nu} \bigr\rangle_{\scr V}
	={}&
	g_{\mu\nu}
	\times
	\boldsymbol{\Gamma}(\nu_{\scr V})
	(1\!-\!\epsilon)^4 H^4
	\biggl[ \Bigl( \frac{D\!-\!3}{2} \Bigr)^{\!2} \!-\! \nu_{\scr V}^2 \biggr]
\nonumber \\
&	\hspace{0.5cm}
	\times 
	\Biggl\{
	- \frac{(D\!-\!1)}{D} \Bigl( \frac{9}{4} \!- \nu_{\scr V}^2 \Bigr) 
	- \frac{3 (D \!-\! 4) (3 \!-\!\epsilon) }{8(1\!-\!\epsilon)} 
	+
	\mathcal{O} \bigl[ (D\!-\!4)^2 \bigr]
	\Biggr\} 
	\, .
\label{Tv primitive}
\end{align}
As before, the two remaining conditions for the counterterm coefficients are most easily 
read off from the~$\bigl( a^2 \delta_\mu^0 \delta_\nu^0 \bigr)$ part, and the most
convenient choice seems to be,
\begin{equation}
\bigl[ \delta Z_\phi \!-\! 6\delta \xi \bigr]_{\scr V}^{\rm div.}
	=
	 \mathcal{O}\bigl[ (D\!-\!4)^0 \bigr]
	 \, ,
\qquad \qquad
\bigl[ \delta \alpha_{\scr R} \bigr]_{\scr V}^{\rm div.} 
	= 
	0
	\, ,
\label{vector counterterms for Tmn A}
\end{equation}
which then uniquely fixes all the divergent parts of the counterterm coefficients tied to vector loops,
\begin{subequations}
\begin{align}
\bigl[ \delta \xi \bigr]_{\scr V}^{\rm div.} 
	={}&
	\frac{ \mu^{D-4} \, \Gamma\bigl( \frac{2-D}{2} \bigr) }{ (4\pi)^{ \frac{D}{2} } }
	\times
	\frac{ ( D \!-\! 2 ) (D \!-\!8) q^2 }{16}
	\, ,
\label{vector counterterm: xi}
\\
\bigl[ \delta \lambda \bigr]_{\scr V}^{\rm div.} 
	={}&
	\frac{ \mu^{D-4} \, \Gamma\bigl( \frac{2-D}{2} \bigr) }{ (4\pi)^{ \frac{D}{2} } }
	\times
	\Bigl[ - (D\!-\!1) q^4  \Bigr] 
	\, ,
\label{vector counterterm: lambda}
\\
\bigl[ \delta Z_{\phi} \bigr]_{\scr V}^{\rm div.} 
	={}&
	\frac{ \mu^{D-4} \, \Gamma\bigl( \frac{2-D}{2} \bigr) }{ (4\pi)^{ \frac{D}{2} } }
	\times
	\biggl[ - \frac{ (D\!-\!1) ( D \!-\! 2 ) q^2 }{2} \biggr]
	\, ,
\label{vector counterterm: Zphi}
\\
\bigl[ \delta \alpha_{\scr R} \bigr]_{\scr V}^{\rm div.} 
	={}&
	\frac{ \mu^{D-4} \, \Gamma\bigl( \frac{2-D}{2} \bigr) }{ (4\pi)^{ \frac{D}{2} } }
	\times
	0
	\, ,
\label{vector counterterm: alphaR}
\end{align}
\end{subequations}
such that $ \bigl[ \delta Z_\phi \!-\! 6\delta\xi \bigr]^{\rm div.} \!\!=\! -7/(32\pi^2)$,
in agreement with~(\ref{vector counterterms for Tmn A}).
Inserting the values for these coefficients in~(\ref{T counterterms})
and adding it to~(\ref{Tv primitive}) produces a finite renormalized result 
upon taking the four-dimensional limit,
\begin{align}
\hspace{-3mm}
\bigl\langle \hat{t}^{\scr \, (2)}_{\mu\nu} \bigr\rangle_{\scr V}
	\! + \bigl[ \delta T_{\mu\nu} \bigr]_{\scr V}^{\rm div.} 
	\!\xrightarrow{D\to4}\!{}&
	- \! \frac{ 3 \bigl( q \overline{\phi} \bigr)^{\!2} }{64 \pi^2 } 
	 H^2
	\Biggl\{
	\Biggl[ 
	(1\!-\!\epsilon)^2 \Bigl( \frac{9}{4} \!-\! \nu_{\scr V}^2 \Bigr) 
		\biggl(\!
		2 \ln\Bigl[ \frac{(1\!-\!\epsilon)H}{\mu} \Bigr]\!
		\!+\! \Psi(\nu_{\scr V})
		\!-\! \frac{1}{2}
		\biggr)
\nonumber \\
&	\hspace{-3.8cm}
	-2 \epsilon
		\biggl( \frac{ (1\!-\!\epsilon)^3 H^2 }{ \bigl( q \overline{\phi} \bigr)^{\!2}  }
		\!-\! \frac{3}{4} \biggr)
	- \frac{(3 \!-\! 4\epsilon) (2 \!+\! 7\epsilon) }{6}
	\Biggr]
	\!\times\! g_{\mu\nu}
	+
	\frac{ 2( 2 \!+\! 7\epsilon ) \epsilon}{3}
	\!\times \! \bigl( a^2 \delta_\mu^0 \delta_\nu^0 \bigr)
	\Biggr\}
	\, .
\label{vector contribution to Tmn 2 ren}
\end{align}
%

\subsubsection{Full one-loop results}
\label{subsubsec: Full one-loop results}

Adding up the scalar loop contribution to the linear~(\ref{scalar contribution to Tmn 1 ren}) 
and quadratic~(\ref{scalar contribution Tmn 2 ren}) parts of the energy-momentum
tensor, and writing them in terms of the scalar perturbation mass~(\ref{scalar mass}) gives,
\begin{align}
\bigl\langle \hat{t}_{\mu\nu} \bigr\rangle_{\scr S}
	={}&
	- 
	\frac{ \lambda }{4 \pi^2}
	\biggl[
	\frac{ 3 \lambda \overline{\phi}^2 \!-\! \bigl( \frac{1}{6} \!-\! \xi \bigr) R }{ 2 \lambda \overline{\phi}^{\,2} } 
	\biggr]
	\Biggl\{
	\biggl[ 
		2 \ln\Bigl[ \frac{(1\!-\!\epsilon)H}{\mu} \Bigr]
		+
		\Psi(\nu_{\scr S})
		+
		\frac{2 \!+\! 15 \epsilon}{6\epsilon} 
\nonumber \\
&	\hspace{2cm}
		- 
		\frac{ ( 1 \!-\! \epsilon ) }{ 2\epsilon ( 1 \!-\! 6 \xi )  }
		+ 
		\frac{ 9 \epsilon (1\!-\!\epsilon) H^2 }{ 2\lambda \overline{\phi}^{\,2} } 
		\biggr]
		\overline{T}_{\mu\nu}
	- 
	\frac{g^{\rho\sigma} \overline{T}_{\rho\sigma } }{8(1\!-\!\epsilon) }
			g_{\mu\nu} 
	\Biggr\}
	\, .
\label{scalar contribution Tmn ren total}
\end{align}
and adding up the vector loop contributions from~(\ref{scalar contribution to Tmn 1 ren}) 
and~(\ref{vector contribution to Tmn 2 ren}), and writing them in terms of scalar
and vector masses~(\ref{scalar mass})  and~(\ref{vector mass}) gives,
\begin{align}
\bigl\langle \hat{t}_{\mu\nu} \bigr\rangle_{\scr V}
	={}&
	- \frac{3 q^2 }{8 \pi^2}
	\Biggl\{
	\biggl[
	\frac{ \bigl( q \overline{\phi} \bigr)^{\!2}
		\!-\! \lambda \overline{\phi}^{\,2} 
		\!+\! \bigl( \frac{1}{6} \!-\! \xi \bigr) R}{ 2 \lambda \overline{\phi}^{\,2} }
	\biggr]
	\biggl[ 
		2 \ln\Bigl[ \frac{(1\!-\!\epsilon)H}{\mu} \Bigr] 
		+ 
		\Psi(\nu_{\scr V})
	+
	\frac{ 1 \!+\! 5 \epsilon }{ 3 \epsilon} 
\nonumber \\
&	\hspace{-1cm}
	- 
	\frac{ ( 1 \!-\! \epsilon ) }{ 3 \epsilon ( 1 \!-\! 6 \xi ) }
	+ 
	\frac{ 3 \epsilon (1\!-\!\epsilon)  H^2 }{ \lambda \overline{\phi}^{\,2} } 
	\biggr] 
	\overline{T}_{\mu\nu}
	+
	\biggl[ 
	\frac{ (2 \!+\! 7\epsilon) }{ 12\epsilon (1\!-\!6\xi) }
	+
	\frac{ 3 \epsilon H^2 }{ 4 \lambda \overline{\phi}^{\,2} } 
	\biggr]
	\overline{T}_{\mu\nu}
\nonumber \\
&
	- 
	\biggl[
	\frac{ \bigl( q \overline{\phi} \bigr)^{\!2}
		\!-\! \lambda \overline{\phi}^{\,2} 
		\!+\! \bigl( \frac{1}{6} \!-\! \xi \bigr) R}{ 2 \lambda \overline{\phi}^{\,2} }
	\biggr]
	\frac{ g^{\rho\sigma} \overline{T}_{\rho\sigma} }{ 8 (1\!-\!\epsilon)} g_{\mu\nu}
	\Biggr\}
	+
	\frac{3 \lambda }{8 \pi^2}
		\!\times\!
		 \frac{ \epsilon (1\!-\!\epsilon)^3 H^4}{ \bigl( \lambda \overline{\phi}^{\,2} \bigr)^{\!2} }
	\overline{T}_{\mu\nu}
	\, ,
\label{vector contribution Tmn ren total}
\end{align}
where~$\overline{T}_{\mu\nu} $
is the tree level result given in~(\ref{scalar classical Tmn 0}).
Recalling that~$\nabla^\mu \overline{T}_{\mu\nu} \!=\! 0$, 
one can easily check 
 that both~(\ref{scalar contribution Tmn ren total})
and~(\ref{vector contribution Tmn ren total}) are separately covariantly 
conserved,~\footnote{Useful relations 
are~$\nabla_\nu \bigl( g^{\rho\sigma} \overline{T}_{\rho\sigma} \bigr) \!=\! 
	-4 \epsilon a H \delta_\nu^0 \bigl( g^{\rho\sigma} \overline{T}_{\rho\sigma} \bigr)$,
and~$\nabla^\mu \bigl[ \ln(a) \overline{T}_{\mu\nu} \bigr] \!=\! - (H/a) \overline{T}_{0\nu}$.} 
as they should
be. We emphasize this is true only after linear and quadratic contributions are added up.
The final expressions~(\ref{scalar contribution Tmn ren total}) and~(\ref{vector contribution Tmn ren total}) 
are rather complicated, and one may be tempted to subtract some
of the terms by a judicious choice of finite counterterms. A closer look at the 
finite part of the energy-momentum tensor counterterms in~(\ref{T counterterms}),
\begin{equation}
\bigl[ \delta T_{\mu\nu} \bigr]^{\rm fin.}
	=
	\biggl\{
		\bigl[ \delta Z_\phi \bigr]^{\rm fin.}
		+
		\frac{ 144 \lambda (2\!-\!\epsilon) }{ (1 \!-\! 6\xi) }
			\frac{H^2}{ \lambda \overline{\phi}^{\,2} } 
			\bigl[ \delta \alpha_{\scr R} \bigr]^{\rm fin.}
	\biggr\}
	\!\times\!
	\overline{T}_{\mu\nu}
	\, .
\end{equation}
where we took account of the simple choice for the condensate correction~(\ref{two finite counters}),
reveals that further simplifications are rather limited if we restrict ourselves to counterterm coefficients
being independent of~$\epsilon$.
Therefore, we shall not pursue this, and instead we shall focus our analysis on
the role of secular effects, and perturbativity of the one-loop 
results~(\ref{scalar contribution Tmn ren total})--(\ref{vector contribution Tmn ren total}).

\section{Various limits}
\label{sec: Various limits}

In this section we discuss various limits of the one-loop
corrections to the condensate~(\ref{varphiS final}) and (\ref{varphiV final}),
and to the energy momentum tensor~(\ref{scalar contribution Tmn ren total}),
and~(\ref{vector contribution Tmn ren total}), and compare them with the results
from the literature when possible.

\subsection{De Sitter limit}
\label{subsec: De Sitter limit}

The de Sitter limit is defined by the constant physical Hubble rate,~$H\!=\!H_0$,
obtained in the limit of vanishing principal slow-roll parameter,~$\epsilon\!\to\!0$,
where the Ricci scalar is~$R_0\!=\!12H_0^2$.
At tree level the condensate~(\ref{scalar amplitude}) is constant, 
while the energy-momentum tensor~(\ref{scalar classical Tmn 0}) vanishes,
\begin{equation}
\overline{\phi} \xrightarrow{\epsilon \to 0} \overline{\phi}_0 
	=
	\pm H_0 \sqrt{\frac{-12\xi}{\lambda}} \, ,
\qquad \qquad
\overline{T}_{\mu\nu}
	\xrightarrow{\epsilon\to0} 0 \, .
\end{equation}
The effective masses of fluctuations are also constant in de Sitter,
\begin{equation}
M_{\scr S}^2 \xrightarrow{ \epsilon \to 0 } m_{\scr S}^2 
	=
	- 24 \xi H_0^2 \, ,
\qquad \quad
M_{\scr V}^2 \xrightarrow{ \epsilon \to 0 } m_{\scr V}^2 
	=
	- \frac{ 12 \xi q^2}{\lambda} H_0^2
	=
	\frac{q^2 m_{\scr S}^2 }{ 2 \lambda }
	\, .
\end{equation}
Consequently, the condensate one-loop corrections~(\ref{varphiS final}) and~(\ref{varphiV final}) are also
constant, and reduce to,
\begin{align}
\frac{ \bigl\langle \hat{\varphi} \bigr\rangle_{\scr S} }{ \overline{\phi}_0 }
	={}&
	\! - 
	\frac{3 \lambda }{16 \pi^2} 
	\biggl[ \frac{ 3 \lambda \overline{\phi}_0^2 - \bigl( \frac{1}{6} \!-\! \xi \bigr) R_0 }
		{ 2 \lambda \overline{\phi}_0^{\,2} } \biggr]
	\biggl[
		2 \ln\Bigl( \frac{H_0}{\mu} \Bigr)
		+
		\Psi \bigl( \nu_{\scr S}^0 \bigr)
	\biggr]
	\, ,
\\
\frac{ \bigl\langle \hat{\varphi} \bigr\rangle_{\scr V} }{ \overline{\phi}_0 }
	={}&
	- \frac{3 q^2 }{16 \pi^2} 
	\biggl[ \frac{ \bigl( q \overline{\phi}_0 \bigr)^{\!2} - \lambda \overline{\phi}_0^2 
		+ \bigl( \frac{1}{6} \!-\! \xi \bigr) R_0 }{ 2 \lambda \overline{\phi}_0^{2} }
		\biggr]
	\biggl[
		2 \ln\Bigl( \frac{H_0 }{\mu} \Bigr)
	+
	\Psi \bigl( \nu_{\scr V}^0 \bigr)
	\biggr]
	\, ,
\end{align}
where the mode function indices are,
\begin{equation}
\nu_{\scr S}^2 \xrightarrow{\epsilon\to0} 
	\bigl( \nu_{\scr S}^0 \bigr)^2
	=
	\frac{9}{4}  - 24\xi
	=
	\frac{9}{4} 
	- \frac{ m_{\scr S}^2 }{H_0^2}
	\, ,
\qquad  \quad
\nu_{\scr V}^2 \xrightarrow{\epsilon\to0} 
	\bigl( \nu_{\scr V}^0 \bigr)^2
	=
	\frac{1}{4}
	- \frac{ 12 q^2 \xi}{\lambda}
	=
	\frac{1}{4} - \frac{ m_{\scr V}^2}{H_0^2}
	\, .
\end{equation}
For some purposes it is more convenient to express the de Sitter limit in terms of 
the effective masses of fluctuations,
\begin{align}
\frac{ \bigl\langle \hat{\varphi} \bigr\rangle_{\scr S} }{ \overline{\phi}_0 }
	\xrightarrow{\epsilon\to0}{}&
	- \frac{3 \lambda }{ 16 \pi^2  }
	\biggl[
	\frac{  m_{\scr S}^2 \!-\! 2H_0^2 }{ m_{\scr S}^2 }
	\biggr]
	\biggl[
		2 \ln\Bigl( \frac{ H_0 }{ \mu } \Bigr)
		+ \Psi\bigl( \nu_{\scr S}^0 \bigr)
		\biggr]
	\, ,
\\
\frac{ \bigl\langle \hat{\varphi} \bigr\rangle_{\scr V} }{ \overline{\phi}_0 }
	\xrightarrow{\epsilon \to 0} {}&
	-
	\frac{3 q^2 }{ 16 \pi^2  }
	\biggl[
	\frac{  m_{\scr V}^2 + 2 H_0^2 }{ m_{\scr S}^2 }
	\biggr]
	\biggl[
	2 \ln\Bigl( \frac{ H_0 }{ \mu } \Bigr)
	+ \Psi\bigl( \nu_{\scr V}^0 \bigr)
	\biggr]
	\, .
\end{align}
Note that no secular corrections remain, as they are proportional to~$\epsilon$ and vanish
in the exact de Sitter limit.
For the energy-momentum tensor we first observe that,
even though the classical 
tree-level contribution vanishes in the de Sitter limit, the
the singular limit, 
\begin{equation}
\frac{1}{\epsilon} \overline{T}_{\mu\nu}
	\xrightarrow{\epsilon\to0}
	- \frac{3 (1\!-\!6\xi)}{4} \bigl( \overline{\phi}_0 H_0 \bigr)^{\!2} g_{\mu\nu}
	\, ,
\end{equation}
is finite,
so that de Sitter limits of one-loop corrections~(\ref{scalar contribution Tmn ren total}) 
and~(\ref{vector contribution Tmn ren total}) are,
\begin{align}
\bigl\langle \hat{t}_{\mu\nu} \bigr\rangle_{\scr S}
	\xrightarrow{\epsilon\to0} {}&
		\frac{1}{128\pi^2} \Bigl( m_{\scr S}^2 - 2 H_0^2 \Bigr)^{\!2} g_{\mu\nu}
		\, ,
\label{TS dS limit}
\\
\bigl\langle \hat{t}_{\mu\nu} \bigr\rangle_{\scr V}
	\xrightarrow{\epsilon\to0} {}&
	 \frac{ 3 }{128 \pi^2} 
		m_{\scr V}^2 \Bigl( m_{\scr V}^2 + 4 H_0^2 \Bigr)
		g_{\mu\nu}
		\, .
\label{TV dS limit}
\end{align}
Curiously, the de Sitter limit of the one-loop energy-momentum tensor exhibits no dependence on the renormalization 
scale~$\mu$, which implies the absence of UV divergences. This is a peculiarity of the nonminimally coupled Abelian Higgs 
model (2.1) that we consider here, where it is the typically
negative non-minimal coupling, satisfying Eq.~(\ref{inequality}),
 that allows for the existence of the 
symmetry-breaking condensate in the attractor regime.

\medskip

One-loop energy momentum tensors of spectator scalars and vectors have been computed in the literature,
but only in the limit of vanishing condensate.
Therefore they cannot be compared directly to our result~(\ref{TS dS limit}) and~(\ref{TS dS limit}),
that comprise of the tadpole contribution from Sec.~\ref{subsec: Part linear in fluctuations}
and the quadratic perturbation contribution of Sec.~\ref{subsec: Part quadratic in fluctuations}.
Each of the contributions are separately conserved in the 
de Sitter limit, as opposed to the general case when only their sum is conserved.
We can only compare the quadratic parts to the existing results in the literature.

Expression~(\ref{scalar perturbations action}), from which the one-loop energy-momentum tensor 
descends from, suggests that
the proper comparison for the scalar loop contribution is comparing the
de Sitter limit of~(\ref{scalar contribution Tmn 2 ren}),
\begin{equation}
\bigl\langle \hat{t}^{\scr \, (2)}_{\mu\nu} \bigr\rangle_{\scr S}
	+
	\bigl[ \delta T_{\mu\nu} \bigr]_{\scr S}^{\rm div.}
	\! \xrightarrow{D\to4}
		\frac{ (1 \!+\! 12\xi) H_0^4 }{32 \pi^2} 
		\Biggl\{ 
			- 36 \xi
			\biggl[ 2 \ln\Bigl( \frac{H_0}{\mu} \Bigr) 
				+ \Psi\bigl( \nu_{\scr S}^0 \bigr) \biggr]
	\! +
	(1 \!+\! 12\xi)
		\Biggr\}
			g_{\mu\nu} 
		\, .
\end{equation}
to the result for the spectator scalar field with non-minimal coupling~$\xi$ and 
mass~$m^2\!=\!3 \lambda \overline{\phi}_0^2 \!=\! - 36 \xi H_0^2$.
This was first computed in~\cite{Dowker:1975tf}, and the reported results agree
with our, up to renormalization scheme ambiguities, and the conformal anomaly.
The de Sitter limit of the vector loop contribution to the quadratic part in~(\ref{quadratic TV}) suggests we should
compare the de Sitter limit of the expectation value~(\ref{vector contribution to Tmn 2 ren}),
\begin{equation}
\bigl\langle \hat{t}^{\scr \, (2)}_{\mu\nu} \bigr\rangle_{\scr V}
	+ \bigl[ \delta T_{\mu\nu} \bigr]_{\scr V}^{\rm div.} 
	\!\xrightarrow{D\to4}
	-\frac{ 3 m_{\scr V}^2 }{64 \pi^2 } 
	\Biggl\{
	\Bigl( m_{\scr V}^2 \!+\! 2 H_0^2 \Bigr) \!
		\biggl[
		2 \ln\Bigl( \frac{H_0}{\mu} \Bigr)
		\!+\! \Psi\bigl( \nu_{\scr V}^0 \bigr)
		\biggr]
	\!-\! \frac{m_{\scr V}^2}{2}
	\!-\! 2H_0^2
	\Biggr\}
	g_{\mu\nu}
	\, . 
\end{equation}
to the Stueckelberg model with mass~$m_{\scr V}^2\!=\! \bigl( q \overline{\phi}_0 \bigr)^{\!2}$ computed 
in~\cite{Belokogne:2015etf,Belokogne:2016dvd}, where the same result was reported, up to renormalization
scheme ambiguities, and the conformal anomaly.

\subsection{Flat space limit}
\label{subsec: Flat space limit}

The Minkowski limit is strictly speaking not accessible naively. However, we can take it by expressing the
results in terms of effective scalar masses first, and then take~$H_0\!\to\!0$ with keeping~$m_{\scr S}$
constant. It is most convenient to do this from the de Sitter limit of the preceding subsection.
Taking into account that digamma functions in reduce to the following forms
in the flat space limit,~\footnote{This follows 
from~$\nu_{\scr S}^0 \,\overset{H_0\to0}{\longsim} \, i m_{\scr S}/H_0$
and~$\nu_{\scr V}^0 \,\overset{H_0\to0}{\longsim} \, i m_{\scr V}/H_0$,
and the limit,
\begin{equation*}
\Psi(i|z|) = 
\psi \Bigl(\frac{1}{2} \!+\! i |z| \Bigr) 
+ \psi\Bigl(\frac{1}{2} \!-\! i |z| \Bigr)
	\ \overset{z \to \infty}{\longsim} \ 
	2 \ln(|z|)
	\, .
\end{equation*}
}
\begin{align}
\Psi\bigl( \nu_{\scr S}^0 \bigr) 
	\ \overset{ H_0\to0 }{ \underset{ m_{\scr S} = {\tt const.} }{\longsim}} \
	2 \ln \Bigl( \frac{m_{\scr S}}{H_0} \Bigr)
	\, ,
\qquad \qquad
\Psi\bigl( \nu_{\scr V}^0 \bigr) 
	\ \overset{ H_0\to0 }{ \underset{ m_{\scr S} = {\tt const.} }{\longsim}} \
	2 \ln \Bigl( \frac{m_{\scr V} }{H_0} \Bigr)
	\, ,
\end{align}
the one-loop tadpole corrections are,
\begin{equation}
\frac{ \bigl\langle \hat{\varphi} \bigr\rangle_{\scr S} }{ \overline{\phi}_0 }
	\xrightarrow[m_{\scr S} = {\tt const.} ]{H_0\to0}
	- \frac{3 \lambda  }{ 8 \pi^2  }
	\ln\Bigl( \frac{ m_{\scr S} }{ \mu } \Bigr)
	\, ,
\qquad
\frac{ \bigl\langle \hat{\varphi} \bigr\rangle_{\scr V} }{ \overline{\phi}_0 }
	\xrightarrow[m_{\scr S} = {\tt const.} ]{H_0\to0}
	- \frac{3 q^2 }{ 8 \pi^2  }
	\frac{ m_{\scr V}^2 }{ m_{\scr S}^2 }
	\ln\Bigl( \frac{ m_{\scr V} }{ \mu } \Bigr)
	\, .
\label{CW tadpoles}
\end{equation}
These correspond to the results of Minkowski space computations~\cite{Irges:2017ztc},
up to renormalization scheme dependent terms.

While we are able to infer the flat space limit from the condensate correction in de Sitter space,
we are not able to do so for the energy-momentum tensor. The reason is that the mass term and the 
non-minimal coupling term are not distinguishable  in de Sitter for the evolution of the scalar.
However, they are distinguished by the energy-momentum tensor which involves the variational derivative
with respect to the metric, which treats the mass term and the non-minimal coupling terms differently.

\subsection{Large field limit/large non-minimal coupling/large masses}
\label{subsec: Large non-minimal coupling/large masses}

 The large field limit~$\overline{\phi}/H \!\gg\!1$ in our model corresponds to the 
limit of large negative nonminimal coupling~$\xi\!\ll\!-1$.\footnote{We still assume the non-minimal coupling
is small enough not to induce significant backreaction on the expansion rate.} 
Here we have for the tree-level condensate~(\ref{scalar amplitude}) and the energy-momentum tensor~(\ref{scalar classical Tmn 0}),
\begin{equation}
\frac{\overline{\phi}}{H}
	\ \overset{\xi \ll -1}{\longsim} \
	\pm \sqrt{ \frac{ - 6 \xi  (2\!-\!\epsilon)  }{\lambda} }
	\, ,
\qquad
\overline{T}_{\mu\nu}
	\ \overset{\xi \ll -1}{\longsim} \
	\frac{ - 3 \xi \epsilon}{2}
	\biggl[
	- ( 3 \!-\! 4\epsilon ) g_{\mu\nu}
	+ 4\epsilon \bigl( a^2 \delta_\mu^0 \delta_\nu^0 \bigr)
	\biggr]
	\bigl( \overline{\phi} H \bigr)^{\!2}
	\, .
\end{equation}
For fluctuations this is essentially the limit of large effective masses,
\begin{equation}
\frac{M_{\scr S}^2}{H^2}
	\ \overset{\xi \ll -1}{\longsim}
	- \! 12 \xi (2 \!-\! \epsilon) \gg 1
\, ,
\qquad \qquad
\frac{M_{\scr V}^2}{H^2}
	\ \overset{\xi \ll -1}{\longsim} 
	- \frac{ 6 \xi  q^2 (2\!-\!\epsilon)  }{\lambda}
	\gg 1 \, ,
\end{equation}
provided the ratio of couplings~$q^2/\lambda$ is not hierarchical. Given that the mode function
indices in this limit reduce to,
\begin{align}
\nu_{\scr S}^2
	\ \overset{\xi \ll -1}{\longsim} \
	\frac{ 12 \xi (2 \!-\! \epsilon) }{ (1\!-\!\epsilon)^2 } 
	\ll -1
	\, ,
\qquad \quad
\nu_{\scr V}^2
	\ \overset{\xi \ll -1}{\longsim} \
	\frac{q^2}{ \lambda } \times \frac{ 6 \xi (2\!-\!\epsilon) }{ (1\!-\!\epsilon)^2 } 
	\ll - 1
	\, ,
\end{align}
which implies that,
\begin{equation}
\Psi(\nu_{\scr S})
	\ \overset{\xi \ll -1}{\longsim} \
	2 \ln \Bigl[ \frac{M_{\scr S} }{ (1\!-\!\epsilon) H } \Bigr]
	\, ,
\qquad \qquad
\Psi(\nu_{\scr V})
	\ \overset{\xi \ll -1}{\longsim} \
	2 \ln \Bigl[ \frac{M_{\scr V} }{ (1\!-\!\epsilon) H } \Bigr]
	\, ,
\end{equation}
and produces the result for the one-loop condensate corrections,
\begin{align}
\frac{ \bigl\langle \hat{\varphi} \bigr\rangle_{\scr S} }{ \overline{\phi} }
	\ \overset{\xi \ll -1}{\longsim}
	- \! \frac{3 \lambda }{ 8 \pi^2  }
	\ln\Bigl( \frac{ M_{\scr S} }{ \mu } \Bigr)
	\, ,
\qquad \qquad
\frac{ \bigl\langle \hat{\varphi} \bigr\rangle_{\scr V} }{ \overline{\phi} }
	\ \overset{\xi \ll -1}{\longsim}
	- \! \frac{3 q^2 }{ 8 \pi^2  }
	\frac{ M_{\scr V}^2}{M_{\scr S}^2}
	\ln\Bigl( \frac{ M_{\scr V} }{ \mu } \Bigr)
	\, .
\end{align}
that are a direct non-equilibrium generalization of the Coleman-Weinberg result~\cite{Coleman:1973jx}
with time-dependent effective masses.

For the one-loop energy-momentum tensor the large field limit is,
\begin{equation}
\bigl\langle \hat{t}_{\mu\nu} \bigr\rangle_{\scr S}
	\ \overset{\xi \ll -1}{\longsim} {}
	- \frac{\lambda}{2\pi^2} 
		\ln\Bigl( \frac{ M_{\scr S} }{ \mu } \Bigr) 
		\!\times\!
		\overline{T}_{\mu\nu}
	\, ,
\qquad
\bigl\langle \hat{t}_{\mu\nu} \bigr\rangle_{\scr V}
	\ \overset{\xi \ll -1}{\longsim} \
	- \frac{ 3 q^2}{4 \pi^2}
		\frac{ M_{\scr V}^2 }{ M_{\scr S}^2}
			\ln\Bigl( \frac{ M_{\scr V} }{\mu} \Bigr)
			\!\times\!
		\overline{T}_{\mu\nu}
			\, ,
\end{equation}
where we neglected constant corrections to the logs (that can nonetheless be 
enhanced by a relative factor~$1/\epsilon$ in comparison when~$\epsilon\!\ll\!1$.)
The expressions above are not conserved on the account of time dependence
of masses of fluctuations. However, the non-conservation is only at the subleading limit 
that is countered by the subleading terms we neglected.

\subsection{Small field limit}
\label{subsec: Small field limit}

The small field limit~$\overline{\phi}/H\!\ll\!1$ is implemented by taking the critical value limit for the 
nonminimal coupling,
\begin{equation}
\xi = \xi_{\rm cr} - \Delta\xi = 
	\frac{\epsilon (3 \!-\! 2\epsilon)}{ 6 (2 \!-\! \epsilon) } - \Delta \xi
	\, ,
\qquad \qquad
0 < \Delta \xi \ll 1 \, ,
\end{equation}
such that the condition~(\ref{inequality}) is close to saturation, where the tree-level condensate
field value is vanishing small,
\begin{equation}
\frac{\lambda \overline{\phi}^{\,2} }{H^2} 
	\ \overset{ \Delta\xi \to 0 }{\longsim} \
	6 (2\!-\!\epsilon) \Delta \xi \, .
\end{equation}
The mode function indices in the small field limit are,
\begin{equation}
\nu_{\scr S}^2 
	= 
	\frac{9}{4} - \frac{ 12 (2\!-\!\epsilon) \Delta\xi }{ (1\!-\!\epsilon)^2 } 
	\, ,
\qquad \qquad
\nu_{\scr V}^2 
	=
	\frac{1}{ 4 }
		- \frac{q^2}{ \lambda } 
			\frac{ 6 (2\!-\!\epsilon) \Delta \xi }{ (1\!-\!\epsilon)^2 }
	\, ,
\end{equation}
so that the digamma functions contribute as,
\begin{equation}
\Psi(\nu_{\scr S})
	\ \overset{ \Delta\xi \to 0 }{\longsim} \
	- \frac{ (1\!-\!\epsilon)^2 }{ 4 (2\!-\!\epsilon) \Delta\xi }
	+ \frac{7}{3} - 2 \gamma_{\scr E}
	\, ,
\qquad
\Psi(\nu_{\scr V})
	\ \overset{ \Delta\xi \to 0 }{\longsim} \
	- \frac{\lambda}{q^2} \frac{ (1\!-\!\epsilon)^2 }{ 6 (2\!-\!\epsilon) \Delta\xi }
	+ 1 - 2 \gamma_{\scr E}
	\, .
\end{equation}
This sufficies to evaluate the limiting behaviour of our results.
The condensate corrections~(\ref{varphiS final}) and~(\ref{varphiS final}) diverge in this limit,
\begin{align}
\frac{ \bigl\langle \hat{\varphi} \bigr\rangle_{\scr S} }{ \overline{\phi} }
	\ \overset{ \Delta\xi \to 0 }{\longsim} {}&
	- 
	\frac{ \lambda(1\!-\!3\epsilon) (1\!-\!\epsilon)^3}{ 128 \pi^2 (2\!-\!\epsilon)^2 \Delta\xi^2 }
	+ \mathcal{O}\bigl( 1/\Delta\xi \bigr)
	\, ,
\\
\frac{ \bigl\langle \hat{\varphi} \bigr\rangle_{\scr V} }{ \overline{\phi} }
	\ \overset{ \Delta\xi \to 0 }{\longsim} {}& \
	\frac{ ( \lambda \!-\! 3\epsilon q^2 ) (1\!-\!\epsilon)^3 }{ 192 \pi^2 (2\!-\!\epsilon)^2 \Delta \xi^2 }
	+ \mathcal{O}\bigl( 1/\Delta\xi \bigr)
	\, ,
\end{align}
which is actually a consequence of the singular definition of the variable. In the unitary gauge we emply
here the scalar field is actually the modulus of the complex scalar field, which is not defined at the origin.
In fact, a closer look at the leading relative condensate correction above reveals it to be negative, meaning
that the meaningful description breaks down before the singular point. 
Similar behaviour is observed for the
energy-momentum tensor corrections,
\begin{align}
\bigl\langle \hat{t}_{\mu\nu} \bigr\rangle_{\scr S}
	\overset{\Delta \xi \to 0}{\longsim} \ {}&
	- \frac{\epsilon (1\!-\!\epsilon)^5 (1\!-\!4\epsilon) }{ 32\pi^2 (2\!-\!\epsilon)^2 \Delta\xi }
	H^4 \Bigl[ - (3\!-\!4\epsilon) g_{\mu\nu} \!+\! 4\epsilon \bigl( a^2 \delta_\mu^0 \delta_\nu^0 \bigr) \Bigr]
	\!+ \mathcal{O} \bigl( \Delta\xi^0 \bigr)
	\, ,
\\
\bigl\langle \hat{t}_{\mu\nu} \bigr\rangle_{\scr V}
	\overset{\Delta \xi \to 0}{\longsim} \ {}&
	\frac{ \epsilon (1\!-\!\epsilon)^5 }{ 32\pi^2 (2\!-\!\epsilon)^2 \Delta\xi  }
	\Bigl[ 1 \!-\! \frac{ 3\epsilon q^2 }{\lambda} \Bigr]
	H^4 \Bigl[ - (3\!-\!4\epsilon) g_{\mu\nu} \!+\! 4\epsilon \bigl( a^2 \delta_\mu^0 \delta_\nu^0 \bigr) \Bigr]
	\! + \mathcal{O} \bigl( \Delta\xi^0 \bigr)
	\, ,
\end{align}
that are finite only in the de Sitter limit,
\begin{equation}
\bigl\langle \hat{t}_{\mu\nu} \bigr\rangle_{\scr S} 
	\xrightarrow[\epsilon\to0]{\Delta \xi \to 0}
	\frac{H^4}{32\pi^2} g_{\mu\nu} 
	\, ,
\qquad \qquad
\bigl\langle \hat{t}_{\mu\nu} \bigr\rangle_{\scr V} 
	\xrightarrow[\epsilon\to0]{\Delta \xi \to 0}
	0
	\, ,
\end{equation}
where the scalar loop contribution reduces to the contribution of the massless, minimally coupled scalar,
supplemented by the finite contribution obtained from taking the singular limit~$\xi\!\to\!0$ of
the nonminimally coupled case (consistent with~\cite{Glavan:2013mra,Glavan:2014uga,Glavan:2015cut}), 
while the vector loop contribution reduces to 
the vanishing one of a massless photon~\cite{Glavan:2022pmk,Glavan:2022dwb,Glavan:2022nrd}.
However, behaviour of the results for nonvanishing~$\epsilon$ points to the need of examining this regime in a different gauge, 
that is adapted to the vanishing condensate
limit, unlike the unitary gauge we use here.

\subsection{Hierarchal couplings}
\label{subsec: Hierarchal couplings}

Interesting limit is provided by a hierarchy between the scalar self-coupling constant
and the~$U(1)$ charge,
\begin{equation}
q^2/\lambda \gg 1 \, ,
\label{hierar}
\end{equation}
when the vector loop provides a parametrically larger contribution
than the scalar one. Furthermore, given that in this limit,
\begin{equation}
\Psi(\nu_{\scr V})
	\ \overset{q^2 \gg \lambda }{\longsim} \
	2 \ln \Bigl[ \frac{q \overline{\phi} }{ (1\!-\!\epsilon) H } \Bigr]
	\, ,
\end{equation}
the dominant contribution to the condensate correction,
\begin{equation}
\frac{ \bigl\langle \hat{\varphi} \bigr\rangle_{\scr V} }{ \overline{\phi} }
	\ \overset{q^2 \gg \lambda }{\longsim}
	- \frac{3 q^4 }{16 \pi^2 \lambda } 
		\ln\Bigl( \frac{ q \overline{\phi} }{\mu} \Bigr)
	\, ,
\end{equation}
and to the energy-momentum tensor correction,
\begin{align}
\bigl\langle \hat{t}_{\mu\nu} \bigr\rangle_{\scr V}
	=
	- \frac{3 q^4 }{8 \pi^2 \lambda }
		\ln\Bigl( \frac{q \overline{\phi} }{\mu} \Bigr) \!\times\! 
		\overline{T}_{\mu\nu}
	\, .
\end{align}
derive from the vector loop.
We see that the hierarchy~(\ref{hierar}) of coupling constants can lead to a big enhancement
of the loop corrections. This is a feature of the model that is present already in flat space. However, the 
secular logarithm multiplying this correction is innate to curved spacetimes, which is discussed in detail
in Sec.~\ref{sec: Late time limit and an RG explanation}.

\subsection{Vanishing~$U(1)$ charge}
\label{subsec: Vanishing U(1) charge}

Even though the limit~$q\!\to\!0$ is a singular limit of the unitary gauge~\cite{Glavan:2020zne},
the scalar field~$\phi$ in the unitary gauge is a gauge-independent observable, and
it is not sensitive to the singularity of the this limit. Therefore, we may infer the
result in the limit of vanishing~$U(1)$ charge. Interestingly, the vector loop contributions do not 
vanish, neither for the condensate,
\begin{align}
\frac{ \bigl\langle \hat{\varphi} \bigr\rangle_{\scr V} }{ \overline{\phi} }
	\xrightarrow{q \to 0} {}&
	\frac{3 \lambda }{16 \pi^2} 
	\!\times\!
	\frac{\epsilon (1\!-\!\epsilon)^3 H^4 }{ \bigl( \lambda \overline{\phi}^{\,2} \bigr)^{\!2} }
	\, ,
\end{align}
nor for the energy-momentum tensor,
\begin{align}
\bigl\langle \hat{t}_{\mu\nu} \bigr\rangle_{\scr V}
	\xrightarrow{q\to0} {}&
	\frac{3 \lambda }{8 \pi^2}
		\!\times\!
		 \frac{ \epsilon (1\!-\!\epsilon)^3 H^4}{ \bigl( \lambda \overline{\phi}^{\,2} \bigr)^{\!2} }
		\!\times\!
	T_{\mu\nu}^{\scr (0)}
	\, .
\end{align}
Nevertheless, inorder to be sure of this limit the computation should be redone in a gauge regular in the
vanishing~$U(1)$ charge limit.

\section{Late time limit and an RG explanation}
\label{sec: Late time limit and an RG explanation}

The time dependence of the Hubble rate in the logarithms of the one-loop corrections
implies the secular breakdown of perturbation theory after long enough time, 
when,
\begin{equation}
\ln \Bigl[ \frac{(1\!-\!\epsilon) H}{\mu} \Bigr]
	\ \overset{a\to\infty}{\longsim} - \epsilon \ln(a) \gg 1
	\, ,
\label{late time log}
\end{equation}
becomes large enough.
This term dominates when~$a\!\gg\! e^{1/\epsilon}$. In inflation this number is very big, but our results
are valid for finite~$\epsilon$. Therefore, the late time limit of the one-loop corrections are,
\begin{equation}
\bigl\langle \hat{\phi} \bigr\rangle =
	\overline{\phi}
	\times
	\Bigl[
	1
	+
	\frac{3 \lambda }{16 \pi^2} 
	C_\phi \times
	\epsilon \ln(a)
	\Bigr]
	\, ,
\qquad
\bigl\langle \hat{T}_{\mu\nu} \bigr\rangle
	=
	\overline{T}_{\mu\nu}
	\times
	\Bigl[
	1
	+
	\frac{3\lambda}{ 8\pi^2 } 
	C_{\scr T} \times \epsilon \ln(a)
	\Bigr]
	\, ,
\label{late time correction}
\end{equation}
where the the scalar and vector loops contribute differently to each constant factor,
\begin{align}
C_\phi ={}&
	\frac{ 3 \lambda \overline{\phi}^2 \!-\! \bigl( \frac{1}{6} \!-\! \xi \bigr) R }
		{ \lambda \overline{\phi}^{\,2} }
	+
	\frac{q^2 }{ \lambda } 
	\biggl[ \frac{ \bigl( q \overline{\phi} \bigr)^{\!2} \!-\! \lambda \overline{\phi}^2 
		\!+\! \bigl( \frac{1}{6} \!-\! \xi \bigr) R}{ \lambda \overline{\phi}^{2} }
		\biggr]
	=
	\frac{q^4}{\lambda^2}
	+
	\frac{1}{\Xi \!-\! 1} \Bigl( \frac{q^2}{\lambda} -1 \Bigr) + 2
		\, ,
\label{Cphi}
\\
C_{\scr T} ={}&
	\frac{ 2 } {3}
	\biggl[
	\frac{ 3 \lambda \overline{\phi}^2 \!-\! \bigl( \frac{1}{6} \!-\! \xi \bigr) R }{ \lambda \overline{\phi}^{\,2} } 
	\biggr]
	\!
	+
	\frac{q^2 }{ \lambda }
	\biggl[
	\frac{ \bigl( q \overline{\phi} \bigr)^{\!2}
		\!-\! \lambda \overline{\phi}^{\,2} 
		\!+\! \bigl( \frac{1}{6} \!-\! \xi \bigr) R}{ \lambda \overline{\phi}^{\,2} }
	\biggr]
	=
	\frac{q^4}{\lambda^2} + \frac{1}{ \Xi \!-\! 1} \Bigl( \frac{q^2}{\lambda} - \frac{2}{3}  \Bigr)
	+ \frac{4}{3}
	\, .
\label{CT}
\end{align}
where~$\Xi\!=\! (1\!-\!6\xi)(2\!-\!\epsilon)/[2(1\!-\!\epsilon)^2]$. 
It is worth noting that 
had we applied the effective potential approximation for our computation,
and was utilized in~\cite{Katuwal:2021kry}, we would have missed the terms of order~$\mathcal{O}(\epsilon)$
in the coefficients above that appear in the dominant late-time contribution. This can be tracked down to 
the classical equation~(\ref{tree-level condensate eom}) that the tree-level condensate is
assumed to satisfy. The effective potential approximation in our case 
would correspond to neglecting the d'Alembertian term in that equation. Even though this drops the time derivatives,
the condensate is still time-dependent because of the non-minimal coupling, and retains the same scaling,
but its amplitude does not account correctly for the~$\mathcal{O}(\epsilon)$ terms.

\begin{figure}[h!]
\centering
\includegraphics[width=7.3cm]{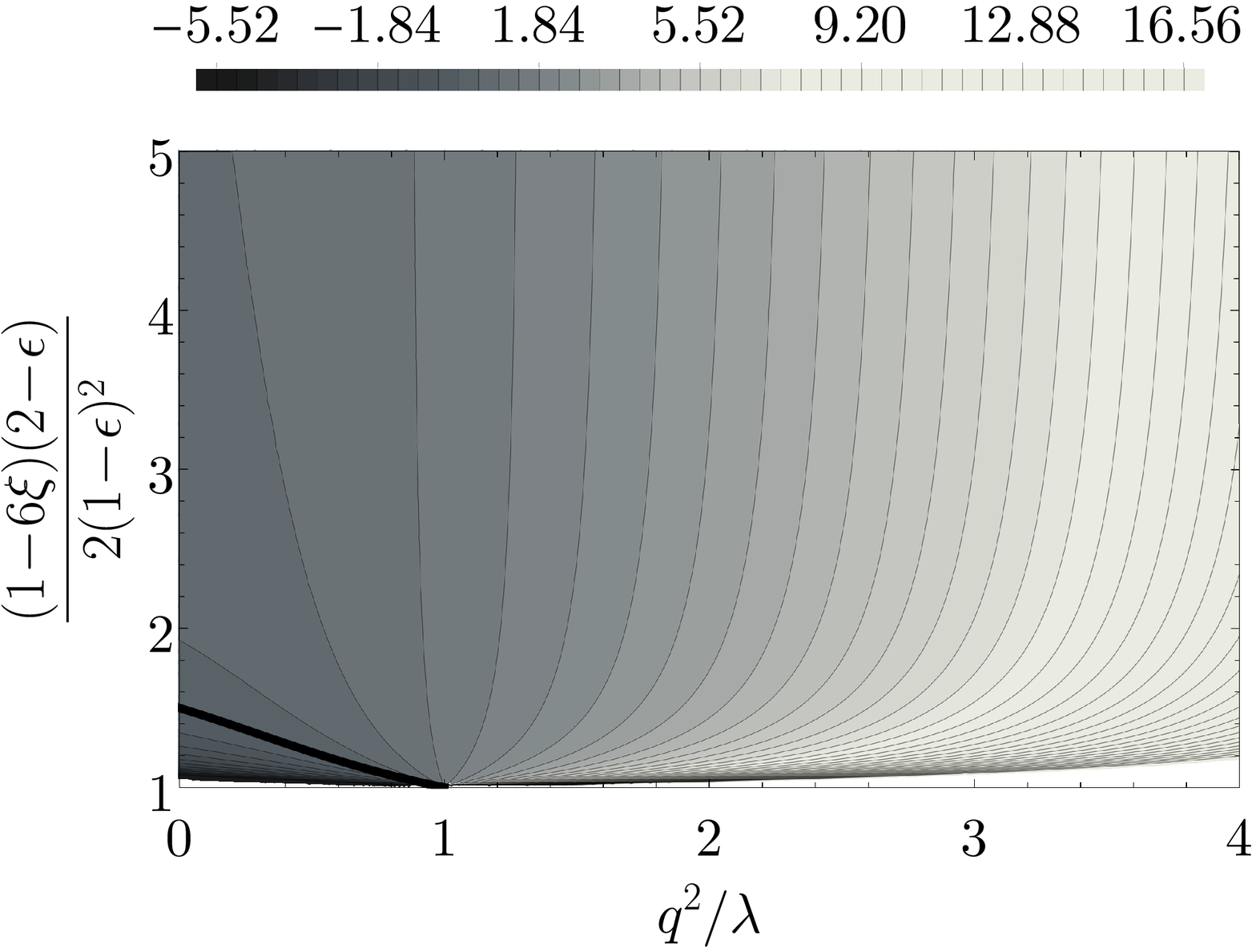}
\hfill
\includegraphics[width=7.3cm]{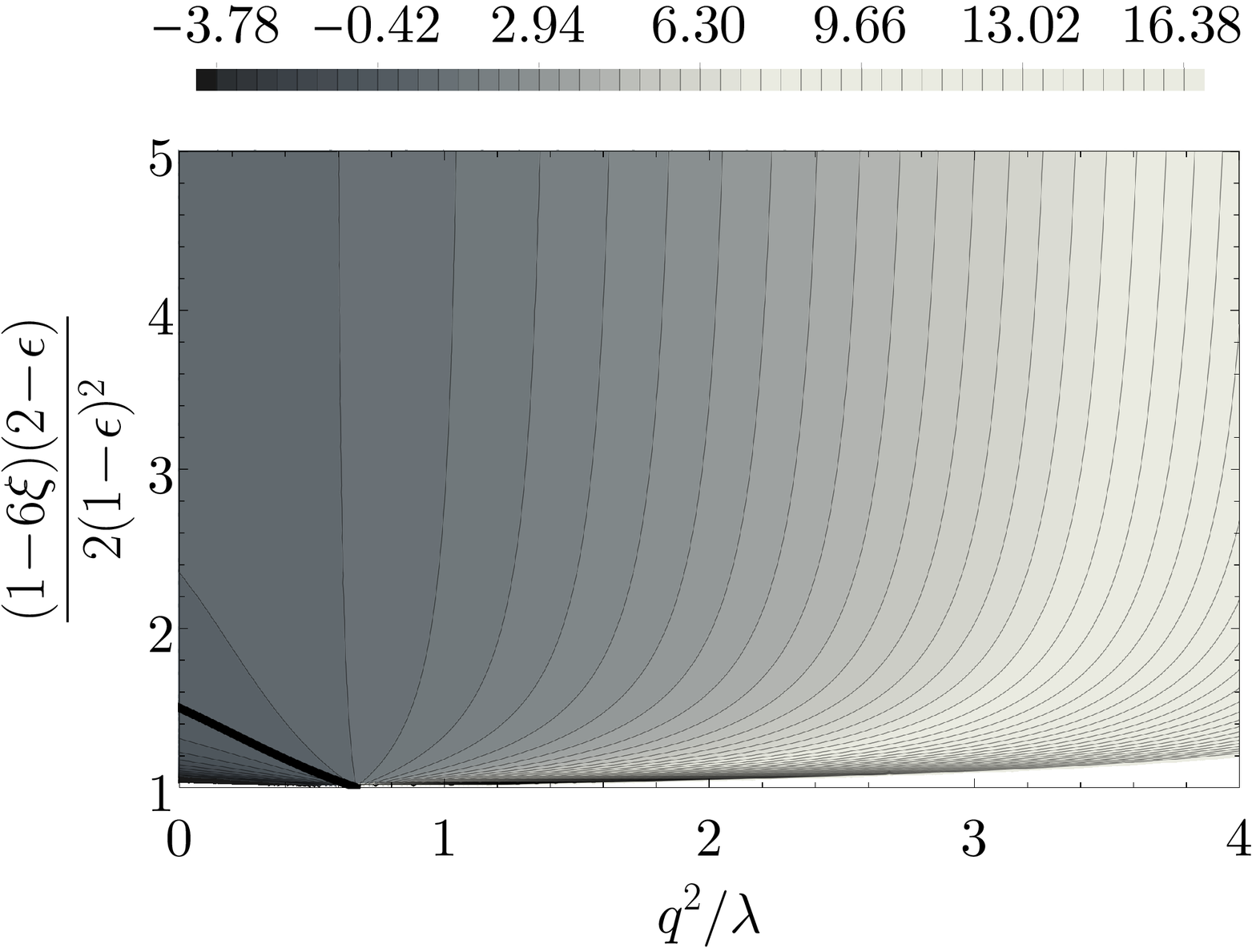}
\vspace{-3mm}
\caption{
Density plots of the coefficient~(\ref{Cphi}) multiplying the late time correction of the condensate ({\it left}),
and of the coefficient~(\ref{CT}) multiplying the late time correction of the energy-momentum tensor ({\it right}).
Thin black curves denote points of equal value of the coefficient; curves are equidistant 
in values
of the coefficients. The value increases from the dark shading to the 
light shading. The thick, black, nearly straight lines denote where the coefficient vanishes,
and the values the coefficients to the left of these lines are negative, while values to the right are positive.}
\label{condensate coefficient}
\end{figure}

The factors~(\ref{Cphi}) and~(\ref{CT}) that multiply the secular corrections in~(\ref{late time correction})
depend only on the two independent combinations of model parameters,~$q^2/\lambda$ and~$\Xi$, 
and this dependence is depicted in Fig.~\ref{condensate coefficient}.
%
The case of hierarchal couplings~$q^2\!\gg\!\lambda$ from Sec.~(\ref{subsec: Hierarchal couplings}) is
particularly interesting. It is captured by a simple expression
independent of the non-minimal coupling,
\begin{equation}
\bigl\langle \hat{\phi} \bigr\rangle 
	\ \overset{a \gg 1}{\longsim} \ 
	\overline{\phi} \biggl[
	1 + \frac{q^4}{\lambda} \frac{3\epsilon}{16 \pi^2} \ln(a)
	\biggr] \, ,
\qquad \qquad
\bigl\langle \hat{T}_{\mu\nu} \bigr\rangle 
	\ \overset{a \gg 1}{\longsim} \ 
	\overline{T}_{\mu\nu}
	\biggl[
	1
	+
	\frac{q^4}{\lambda} 
	 \frac{3\epsilon}{8 \pi^2} \ln(a)
	\biggr] \, ,
\end{equation}
where the hierarchy of coupling constants 
 can make up for the~$\epsilon$-suppression in inflation.

\medskip

Looking at the late time-expressions~(\ref{Cphi}), in particular the condensate correction on the left,
it might be tempting to resum the secular
correction,~$\bigl\langle \hat{\varphi} \bigr\rangle \!\to\! \overline{\phi}_0 a^{-\epsilon(1-\delta)}$
into a renormalized slow-roll parameter~$\epsilon\!\to\!\epsilon(1 \!-\! \delta)$,
where~$\delta \!=\! 3\lambda C_\phi/(16\pi^2)$. This resummation would be correct provided that
the equation of motion for~$\bigl\langle \hat{\varphi} \bigr\rangle$ is linear and autonomous, and that the
secular logarithm is an artifact of the time-dependent perturbation theory. This is very often
what is implicitly or explicitly assumed when the so-called Dynamical renormalization group is invoked to
resum the secular corrections (see e.g.~\cite{Burgess:2009bs,Kamenshchik:2020yyn}). However, the original 
construction of the Dynamical renormalization group~\cite{Chen:1995ena} 
is essentially a reformulation of multiple scale analysis, and consequently necessitates knowing 
the dynamical equation before anything can be resummed, and not only the first few 
perturbative corrections. The fact that this resummation does not 
work here is obviated by the energy-momentum correction that cannot be resummed by the same
slow-roll parameter correction on the account of the coefficients in~(\ref{Cphi}) and~(\ref{CT}) being 
different.
In fact, the secular correction in~(\ref{late time correction}) is an effect of ultraviolet physics
in a nonequilibrium setting. The appropriate framework
 to describe (and eventually resum) them
is the standard renormalization group formalism, implemented in a nonequilibrium setting.
The following section is devoted to discussing some aspects of it.

\subsection{Renormalization group explanation}
\label{subsec: Renormalization group explanation}

The secular corrections in~(\ref{late time correction}) at late times dominate over the tree-level result,
and thus invalidate the perturbative approach. The first step towards extending the results past this point
is to understand which equation governs these secular corrections. It is that equation that ultimately provides 
the resummation scheme. Here we demonstrate that, at the perturbative level, the secular corrections
are governed by the standard renormalization group equations applied to a nonequilibrium setting.

The utility of the renormalization group hinges on our ability to recognize the relevant 
{\it reference scale}~$\mu_0$ that appears together with the arbitrary renormalization scale~$\mu$,
as a logarithm of a dimensionless ratio,~$ \ln(\mu_0/\mu)$.
The appropriate choice of the reference scale ultimately allows 
the renormalization group formalism to effectively resum all such logarithms,
thus extending the validity of perturbative results.
 For systems with multiple
physical scales, such as the one we consider here, different regimes will have different reference
scales~\cite{Elizalde:1993qh}. 
Moreover, for dynamical systems the reference scale will in general be {\it time-dependent}.
Looking at the final one-loop corrections~(\ref{varphiS final})--(\ref{varphiV final}) for the condensate,
and~(\ref{scalar contribution Tmn ren total})--(\ref{vector contribution Tmn ren total}) for the 
energy-momentum tensor, the natural choice for the reference scale
seems to be~$\mu_0 \!=\! (1\!-\!\epsilon)H$. However, as various
limits considered in Sec.~\ref{sec: Various limits} demonstrate, the relevant reference scale depends 
on the parameter
regime. It is the digamma functions, that are inextricably tied to the logarithms
they appear with, that set the correct 
reference scale, without having to choose it by hand. This is perhaps most evident in the hierarchal limit
of Sec.~\ref{subsec: Hierarchal couplings}, 
where the reference scale is recognized as~$\mu_0\!=\! q \overline{\phi}$.

The system at hand is a multiscale problem, for which in general it is not possible to choose a single
reference scale which captures the large logarithms in all regimes.
Here this is essentially due to corrections descending from two independent loops
--- the scalar and the vector one. Potentially large logarithms are captured by the reference
scale if it accounts for the digamma function accompanying the logarithm. In our case we can 
make two choices
which capture either the full large logarithms descending from the scalar loop,
or ones descending from the vector loop, respectively,
\begin{equation}
\ln\Bigl[ \frac{(1\!-\!\epsilon)^2 H^2}{\mu^2} \Bigr] + \Psi(\nu_{\scr S}) 
	\longrightarrow 
	\ln\Bigl( \frac{\mu_0^2}{\mu^2} \Bigr)
	\, ,
	\quad
	{\tt or}
\quad
\ln\Bigl[ \frac{(1\!-\!\epsilon)^2 H^2}{\mu} \Bigr] + \Psi(\nu_{\scr V}) 
	\longrightarrow 
	\ln\Bigl( \frac{\mu_0^2}{\mu^2} \Bigr)
	\, ,
\end{equation}
where~$(1\!-\!\epsilon)^2H^2 \!=\! \frac{1}{2} \big[ \big(\frac{1}{6}\!-\!\xi\big)R -\lambda\overline\phi^{\,2} \big] \!>\! 0$.
 However, we cannot do both simultaneously.
That would require utilizing the full machinery of multiscale renormalization 
group~\cite{Einhorn:1983fc,Ford:1994dt,Ford:1996yc,Ford:1996hd}.
Nevertheless, this is not necessary for our purposes, as the late time behaviour is not marred by such subtleties. In fact,
there is universality in
the late-time secular corrections~(\ref{late time correction}).
This is because the tree-level quantities that make up different
reference scales all scale like the Hubble rate. Therefore, whatever~$\mu_0$ precisley is, at late times 
we always have,
\begin{equation}
\ln \Bigl( \frac{\mu_0}{\mu} \Bigr)
	\ \overset{a\to\infty}{\longsim} - \epsilon \ln(a)
	   \ll -1
	 \, ,
\label{late time s}
\end{equation}
even if~$m_{\scr S} \!\gg\! H$ or~$m_{\scr V}\!\gg\! H$.

The observations above can be formalized using the renormalization group machinery.
Our results depend on the arbitrary renormalization scale~$\mu$,
\begin{equation}
\bigl\langle \hat{\phi} \bigr\rangle
	= 
	\bigl\langle \hat{\phi} \bigr\rangle \Bigl(
		H, \bigl\{ \lambda_n \bigr\} , \ln\bigl[ \tfrac{ \mu_0 }{ \mu } \bigr]
		\Bigr)
		\, ,
\qquad \quad
\bigl\langle \hat{T}_{\mu\nu} \bigr\rangle
	= 
	\bigl\langle \hat{T}_{\mu\nu} \bigr\rangle \Bigl(
		H, \bigl\{ \lambda_n \bigr\}  , \ln\bigl[ \tfrac{ \mu_0 }{ \mu } \bigr]
		\Bigr)
		\, ,
\end{equation}
where~$\bigl\{ \lambda_n \bigr\} \!=\! \{ \lambda, \xi, Z_\phi, \alpha_{\scr R} \}$
stands for all the coupling constants.
Note that we have chosen to treat the
scalar wavefunction renormalization as a coupling constant with its associated~$\beta$-function
for convenience, instead of treating it as 
the anomalous dimension $\gamma$ of the field; this
distinction is essentially immaterial~\cite{Coleman:1973jx}.
The renormalization scale~$\mu$ does not have a physical meaning by itself. What does have physical sense 
is performing several measurements at some scale~$\mu_*$. This process determines the couplings as 
functions of the ratio~$\mu/\mu_*$
\begin{equation}
\lambda_n \rightarrow \lambda_n \bigl( \ln\bigl[ \tfrac{\mu_*}{\mu} \bigr] \bigr) \, ,
\end{equation}
such that the dependence on~$\mu$ completely disappears from the
result,
\begin{align}
&
\mu \frac{d}{d\mu} 
	\bigl\langle \hat{\phi} \bigr\rangle \Bigl(
		H, 
		\bigl\{ \lambda_n\bigl( \ln\bigl[ \tfrac{\mu_*}{\mu} \bigr] \bigr) \bigr\} ,
		\ln\bigl[ \tfrac{ \mu_0 }{ \mu } \bigr]
		\Bigr) = 0 \, ,
\\
&
\mu \frac{d}{d\mu} \bigl\langle \hat{T}_{\mu\nu} \bigr\rangle \Bigl(
		H, 
		\bigl\{ \lambda_n\bigl( \ln\bigl[ \tfrac{\mu_*}{\mu} \bigr] \bigr) \bigr\} , 
		\ln\bigl[ \tfrac{ \mu_0 }{ \mu } \bigr]
		\Bigr) = 0 \, ,
\end{align}
which is guaranteed by the running couplings satisfying the running equations,
\begin{equation}
\mu \frac{d }{d\mu}  \lambda_n = \beta_n \bigl( \{ \lambda_m \} \bigr) \, ,
\label{RG equations}
\end{equation}
where~$\beta_n$ are the~$\beta$-functions associated to couplings~$\lambda_n$, that are determined by the
structure of divergences.
The one-loop~$\beta$-functions
are obtained by acting $-\mu \frac{\partial}{\partial \mu}$
on the counterterm coefficients
given in sections~\ref{subsubsec: Scalar loop contribution}
and~\ref{subsubsec: Vector loop contribution}, 
and taking the limit $D \!\rightarrow \!4$,
\begin{equation}
\beta_\xi
	=
	-\frac{ \lambda (1\!-\!6\xi) }{16 \pi^2}
     + \frac{ q^2 }{16 \pi^2}
	\, ,
\quad \
\beta_\lambda
	=
	 \frac{9 \lambda^2 }{ 8 \pi^2 }
	 +\frac{ 3 q^4 }{ 8 \pi^2 }
	\, ,
\quad \
	\beta_{Z_\phi}
	=
	 \frac{ 3 q^2 }{ 8 \pi^2 }
	\, ,
\quad \
\beta_{{\alpha}_{\scr R}}
	=
	-\frac{ (1\!-\!6\xi)^2 }{1152 \pi^2}
\, .
\label{leading order beta functions}
\end{equation}
It is straightforward to check that our one-loop corrections satisfy the Callan-Symanzik equations to the
given order,
\begin{equation}
\mu \frac{\partial}{\partial\mu} \bigl\langle \hat{\varphi} \bigr\rangle
	=
	- \beta_n \frac{\partial}{\partial \lambda_n} \overline{\phi}
	\, ,
\qquad \qquad
\mu \frac{\partial}{\partial\mu} \bigl\langle \hat{t}_{\mu\nu} \bigr\rangle
	=
	- \beta_n \frac{\partial}{\partial \lambda_n} \overline{T}_{\mu\nu}
	\, ,
\end{equation}
where we have reintroduced an arbitrary~$Z_\phi$
and~$\alpha_{\scr R}$ dependence in the tree-level expressions,
\begin{align}
\overline{\phi} \bigl( H,  \xi, \lambda, Z_\phi \bigr) ={}&
	\pm H \sqrt{ \frac{1}{\lambda} \Bigl[ Z_\phi \epsilon (3\!-\!2\epsilon) - 6\xi(2\!-\!\epsilon) \Bigr] } 
	\, ,
\label{tree-level phi}
\\
\overline{T}_{\mu\nu} \bigl( H, \xi, \lambda, Z_\phi, \alpha_{\scr R} \bigr) ={}&
	\biggl[
	-\frac14 ( 3 \!-\! 4\epsilon ) \epsilon g_{\mu\nu}
	+ \epsilon^2 \bigl( a^2 \delta_\mu^0 \delta_\nu^0 \bigr)
	\biggr]
\nonumber \\
&	\hspace{2cm}
	\times
	\biggl[
	 (Z_\phi \!-\!6\xi)
	\bigl( \overline{\phi} H \bigr)^{\!2}
	+
	144 \alpha_R  (2\!-\!\epsilon) H^4 
	\biggr]
	\, .
\label{tree-level T}
\end{align}
Then, having correctly identified the relevant late-time reference scale, as described at the beginning of this section,
we immediately infer the late-time limit,
\begin{equation}
\bigl\langle \hat{\varphi} \bigr\rangle
	\ \overset{a \to \infty}{\longsim} \
	\epsilon \ln(a) \times \beta_n \frac{\partial}{\partial \lambda_n} \overline{\phi}
	\, ,
\qquad \qquad
\bigl\langle \hat{t}_{\mu\nu} \bigr\rangle
	\ \overset{a \to \infty}{\longsim} \
	\epsilon \ln(a) \times \beta_n \frac{\partial}{\partial \lambda_n} \overline{T}_{\mu\nu}
	\, .
\end{equation}

Apart from efficiently reproducing the late-time correction obtained from the loop computation, 
the real power of the renormalization
group is in its ability to resum the large logarithms. This is accomplished
by first making a variable substitution, and adopting the dimensionless ratio
that includes the reference scale~$\mu_0$ as the new variable,
\begin{equation}
s = \ln\Bigl( \frac{\mu_0}{\mu} \Bigr) \, ,
\end{equation}
such that the Callan-Symanzik equations read,
\begin{equation}
\frac{d}{ds}
	\bigl\langle \hat{\phi} \bigr\rangle \Bigl(
		H, \lambda_n\bigl( \ln\bigl[ \tfrac{\mu_*}{\mu_0} \bigr] \!+\! s \bigr) , s
		\Bigr) = 0 \, ,
\qquad
\frac{d}{ds} \bigl\langle \hat{T}_{\mu\nu} \bigr\rangle \Bigl(
		H, \lambda_n\bigl( \ln\bigl[ \tfrac{\mu_*}{\mu_0} \bigr] \!+\! s \bigr) , s
		\Bigr) = 0 \, ,
\label{s independence of 1 point functions: 1}
\end{equation}
and the equations for the running couplings,
\begin{equation}
\frac{d}{ds} \lambda_n = \beta_n \bigl( \{ \lambda_m \} \bigr) \, .
\label{coupling RG eqs}
\end{equation}
Once we have solved for the running couplings as functions of~$s$, the Callan-Symanzik equations
allow us to express the results choosing any~$s$. Particularly useful is the choice
that removes as much of the explicit~$s$-dependence from the results, and translates it to the~$s$-dependence
of the couplings. This choice corresponds to the following,
\begin{equation}
\bigl\langle \hat{\phi} \bigr\rangle 
=
\bigl\langle \hat{\phi} \bigr\rangle 
	\Bigl(
		H, \lambda_n\bigl( \ln\bigl[ \tfrac{\mu_*}{\mu_0} \bigr] \bigr) , 0
		\Bigr)
	\, ,
\qquad \quad
\bigl\langle \hat{T}_{\mu\nu} \bigr\rangle 
=
\bigl\langle \hat{T}_{\mu\nu} \bigr\rangle 
	\Bigl(
		H, \lambda_n\bigl( \ln\bigl[ \tfrac{\mu_*}{\mu_0} \bigr] \bigr) , 0
		\Bigr)
		\, .
\label{s independence of 1 point functions: 2}
\end{equation}
provided we had identified the relevant reference scale~$\mu_0$.
 What~(\ref{s independence of 1 point functions: 2}) accomplishes when compared 
with~(\ref{s independence of 1 point functions: 1}) 
is that the explicit dependence on the Hubble rate {\it via} $s$ is moved to the coupling constants.
Working at the one-loop level corresponds to solving Eqs.~(\ref{coupling RG eqs}) 
to linear order in~$s$,
\begin{subequations}
\begin{align}
\xi(s) ={}&
	\xi(0)
	- \biggl[ \frac{ \lambda(0) \big[ 1 \!-\! 6\xi(0)  \big] }{ 16 \pi^2 }
		- \frac{ q^2(0) }{ 16 \pi^2 } \biggr] s
		\, ,
\\
\lambda(s) ={}&
	\lambda(0)
	+ \biggl[ \frac{9 \lambda^2(0) }{ 8 \pi^2 }
		+ \frac{ 3 \, q^4(0) }{ 8 \pi^2 } \biggr] s
		\, ,
\\
Z_\phi(s) ={}&
	Z_\phi(0)
	+
	\frac{ 3 \, q^2(0) }{ 8 \pi^2 } s
	\, ,
\\
\alpha_{\scr R}(s) ={}&
	\alpha_{\scr R}(0)
	-
	\frac{ \big[ 1\!-\!6 \xi(0) \big]^2 }{1152 \pi^2} s
\,.
\end{align}
\end{subequations}
It is straightforward to see that when these solutions are plugged into the tree-level result they reproduce
exactly the late-time behaviour,
\begin{align}
\bigl\langle \hat{\phi} \bigr\rangle
	={}& 
	\overline{\phi} 
	+ \bigl\langle \hat{\varphi} \bigr\rangle
	\equiv \overline{\phi} \bigl( H, \xi(s), \lambda(s), Z_\phi(s) \bigr)
	\, ,
\label{running phi}
\\
\bigl\langle \hat{T}_{\mu\nu} \bigr\rangle
	={}& 
	\overline{T}_{\mu\nu}
	+ \bigl\langle \hat{t}_{\mu\nu} \bigr\rangle
	\equiv 
	\overline{T}_{\mu\nu} \bigl( H, \xi(s), \lambda(s), Z_\phi(s), \alpha_{\scr R}(s) \bigr)
	\, .
\label{running T}
\end{align}
Note that, even though all of the time-dependence has been absorbed into effective
running of the coupling constants, this does not imply that the effect is not observable.
On the contrary, quantum corrections will cause the ratio~$\bigl\langle \hat{\phi} \bigr\rangle/H$
to acquire time dependence, which is a physical effect.
We emphasize that the classical form in~(\ref{running phi})--(\ref{running T})
is in general achieved only at asymptotically late times. At intermediate times, 
other logarithmically enhanced contributions can in general occur. 
However, since we are here primarily interested in the question
of restoring perturbativity at late times, this analysis suffices.

The analysis of this section opens up an exciting possibility of resumming the corrections past the 
breakdown of the perturbative expansion, and obtaining reliable late-time behaviour 
utilizing the machinery of the renormalization group. Accomplishing this requires
supplementing the equations for the running couplings ~(\ref{coupling RG eqs})
with an additional two for the~$U(1)$ charge~$q$, and for the vector wavefunction 
renormalization~$Z_{\scr A}$, and then solving the entire system exactly. 
For this we first need the one-loop $\beta$-functions for~$q$ and~$Z_{\scr A}$ in the unitary gauge, 
which do not appear at the level of one-loop analysis we performed here,
but are important for resumming the late-time behaviour correctly. This is left for future work.
Finally, we emphasize that the renormalization group can explain the large
ultraviolet logarithms, but not the infrared logarithms~\cite{Woodard:2008yt},
which in inflation are captured by a 
variant of Starobinsky's stochastic formalism~\cite{Starobinsky:1986fx,Starobinsky:1994bd}, 
as was accomplished in~\cite{Miao:2021gic,Glavan:2021adm}.


\section{Discussion}
\label{sec: Discussion}

Quantum effects in inflation are widely studied utilizing de Sitter space as
an idealized model space for inflation, and
quantum loop corrections have been calculated in many models of interest.
On the other hand, comparatively little is known about quantum effects in more realistic 
models of inflation, where the expansion rate is an adiabatic function of time.
Of particular interest are models which exhibit novel secular effects that do not occur
in de Sitter limit, and this work is devoted to one such system.
We consider power-law inflation, characterized by a constant
principal slow-roll parameter~$\epsilon\!=\!-\dot{H}/H^2$, as a mathematically tractable model
of inflating background,
on which we studied the behaviour of a spectator non-minimally coupled Abelian Higgs model.
The Abelian Higgs model is considered to be in a classical attractor regime,
characterized by the scaling behaviour,~$\overline{\phi}/H\!=\!{\rm const.}$, 
in which the condensate tracks the evolution of the Hubble 
rate. 
This attractor can be seen as a dynamical generalization of 
a symmetry-breaking minimum.
We computed dimensionally regulated one-loop corrections to 
the condensate depicted in Fig.~\ref{tadpole diagrams},
and to the energy-momentum tensor depicted in Figs.~\ref{graviton tadpole 1}
and~\ref{graviton tadpole 2}.
The tadpole contributions to the energy-momentum tensor in Fig.~\ref{graviton tadpole 1}
are often not considered, but are of essential importance because, only when they are added
to the contributions from vacuum fluctuations in Fig.~\ref{graviton tadpole 2},
is the one-loop energy-momentum tensor conserved.
The computations are performed in the unitary gauge,\footnote{
We find no obstacle in computing dimensionally regulated loop correction in
the unitary gauge, conforming with the experience from flat space equilibrium 
computations~\cite{Irges:2017ztc}. 
However, Ref.~\cite{Mooij:2011fi} reports problems with the unitary gauge for dynamical 
scalar condensates, when compared to the covariant gauge computations.
This issue should be resolved by repeating the explicit computations of this paper in 
some analogue of the~$R_\xi$ gauge.
}
in which
both quantities of our interest are manifestly gauge-independent observables.
Our results~(\ref{varphiS final})--(\ref{varphiV final})
and~(\ref{scalar contribution Tmn ren total})--(\ref{vector contribution Tmn ren total})
capture both the infrared and ultraviolet effects,
and they go beyond the effective potential approximation, 
in that they descend from
the full one-loop effective action corrections, which include kinetic term 
corrections that need not be small for finite~$\epsilon$ backgrounds.

Particularly interesting are the secular corrections to both the condensate and the energy-momentum tensor
given in~(\ref{late time correction})--(\ref{CT}), 
which dominate at late times. 
The corrected late time result takes the form of the tree-level result multiplied by a time-dependent amplitude.
Since this amplitude grows in time, the corrections can be seen as driving the evolution away from the classical attractor 
given in~(\ref{scalar condensate}) and~(\ref{scalar amplitude}), pointing
to the quantum instability of the attractor.
Classically the condensate rolls down the potential
tracking the evolution of the Hubble rate. The late time one-loop correction is enhanced by a logarithm of the
scale factor (the number of e-foldings) multiplied by the coefficient that takes positive values for the most
range of the model parameters, as seen from Fig.~\ref{condensate coefficient}. Thus the one-loop correction 
tends to slow down the rolling of the spectator scalar down its potential.

It is interesting that the secular corrections we find are multiplied by the principal slow-roll 
parameter~$\epsilon$,
and thus are suppressed in inflation, and
vanish in the exact de Sitter limit~$\epsilon\!=\!0$, that is often employed as a tractable model
of inflation. This points to the fact that power-law inflation, and other classes of inflating
spacetimes closer to more realistic slow-roll inflation can harbour {\it qualitatively} different, 
and potentially important effects. 
While in inflation the slow-roll parameter acts as a suppression factor, our results are
valid for finite constant (or adiabatically evolving)~$\epsilon$, including instances where it does not
act as a suppression. Nonetheless, given enough time the secular correction will necessarily overcome 
the possible suppressions coming from the slow-roll parameter or the coupling constants,
leading to the breakdown of perturbation theory.
Similar types of corections have also been noticed~\footnote{
It is interesting that for quantum-gravitational corrections found in~\cite{Frob:2018tdd,Lima:2020prb}
the one-loop correction is actually decaying with respect to the tree-level result. Even though there
is a~$\ln(a)$ enhancement factor descending from the UV corrections, just like the one we find here,
the correction decays as a power law as it is additionally multiplied by a decaying~$H^2$ that
has to be there to form a dimensionless product with the dimensionful loop-counting 
parameter~$\kappa^2\!=\! 16 \pi G_{\scr N}$, where $G_N$ is the Newton's constant.} 
in~\cite{Janssen:2009pb,Frob:2018tdd,Lima:2020prb,Katuwal:2021kry}.
This serves as a very good motivation for further studies of quantum corrections
to observables in power-law inflation, 
using both perturbative~\cite{Janssen:2009pb,Glavan:2019uni},
and non-perturbative~\cite{Prokopec:2015owa,Cho:2015pwa,Markkanen:2017edu} methods.

Another observation to be made is that the possible hierarchy between the coupling 
constants,~$q^2 \!\gg\!\lambda$, discussed in Sec.~\ref{subsec: Hierarchal couplings},
leads to an unsuppressed loop correction. This correction is furthermore made bigger at late times
by the secular logarithm, exacerbating the problem of applying perturbation theory. 
Unlike the secular correction, the coupling hierarchy enhancement survives in the de Sitter limit.
In fact it is already seen in flat space~\cite{Coleman:1973jx}, 
where the vector loop moves the minimum of the 
effective potential by an amount larger than tree-level. Despite that one expects 
perturbativity to be respected at two- and higher loops~\cite{Ford:1992pn,Nielsen:2022oec}.
In cosmological expanding spaces, on the other hand, this perturbativity will necessarily 
be spoiled at late times due to secular corrections we report here.

The secular corrections we find are an ultraviolet effect, and can ultimately be accounted for by the
renormalization group, as we show in Sec.~\ref{subsec: Renormalization group explanation}. This is
because the secular logarithm always appears together with the logarithm of the renormalization scale~$\mu$
in the form~$\ln\bigl[ (1\!-\!\epsilon)H / \mu \bigr]$ given in~(\ref{late time log}). Consequently, the secular
correction can be captured by an effective
time-dependent running of the coupling constants given by the 
renormalization group equations. 
Note that this does not imply that all the physical scales are simply rescaled
differently at different points in time. It is the fact that the quantum-corrected condensate no longer scales
as the Hubble rate, and it is their ratio that has a physical meaning. It would be interesting to investigate 
the full power of the renormalization group to 
resum the late time behaviour, and provide reliable results beyond 
the breakdown of perturbation theory.
For the case at hand, that would first require computing the beta functions for the~$U(1)$ charge and the vector field wavefunction renormalization, 
which do not appear in the strictly perturbative computation at hand.
Early efforts in this direction are undertaken in Refs.~\cite{Miao:2021gic,Glavan:2021adm},
but a lot more is needed to fully develop the formalism, especially 
for its higher loop implementation.

It would, furthermore, be interesting to examine 
quantum corrections to the condensate when
the scalar field is the inflaton~\cite{Janssen:2008dw}
as well as the quantum backreaction from cosmological 
perturbations~\cite{Mukhanov:1996ak,Abramo:1997hu,Abramo:1998hi,Abramo:1998hj,Abramo:2001dc}.
A notable model is Higgs inflation~\cite{Bezrukov:2007ep,Bezrukov:2013fka}, and 
stability of the quantum model is of particular interest,
which can be addressed without~\cite{Bezrukov:2014ipa}, or with taking into account 
the background curvature 
corrections~\cite{Herranen:2014cua,Herranen:2015ima,Figueroa:2017slm}
and additional fine tuning problems~\cite{Miao:2015oba,Liao:2018sci,Miao:2019bnq,Miao:2020zeh,Kyriazis:2019xgj,Sivasankaran:2020dzp,Katuwal:2021kry}.
Important questions to be addressed are how 
the secular corrections discussed in this work affect stability of
Higgs inflationary models, as well as the amplitude of the scalar and tensor spectra.
Secular corrections to the reheating period following Higgs inflation is also of great
 interest~\cite{Herranen:2015ima,Figueroa:2017slm,Ema:2016dny,Katuwal:2022szw}.

\section*{Acknowledgements}

We thank Arttu Rajantie and Markus Fr\"ob for discussions on the topic.
DG was supported by the European Union and the Czech Ministry of Education, 
Youth and Sports 
(Project: MSCA Fellowship CZ FZU I --- CZ.02.01.01/00/22\textunderscore010/0002906).
This work is part of the Delta ITP consortium, a program of the Netherlands Organisation
for Scientific Research (NWO) that is funded by the Dutch Ministry of Education, Culture
and Science (OCW) --- NWO project number 24.001.027.



\end{document}